\documentclass{mn2e}
\usepackage{graphicx}

\begin{document}
\title{Population synthesis of wide binary millisecond pulsars}
\author[B. Willems and U. Kolb]
{B. Willems\thanks{E-mail: B.Willems@open.ac.uk, U.C.Kolb@open.ac.uk}
and U. Kolb$^\star$ \\
Department of Physics and Astronomy, The Open University,
Walton Hall, Milton Keynes, MK7 6AA, UK}

\date{Accepted ... Received ...; in original form ...}

\pagerange{\pageref{firstpage}--\pageref{lastpage}} \pubyear{2002}

\maketitle

\label{firstpage}

\begin{abstract}
Four evolutionary channels leading to the formation of wide binary
millisecond pulsars are investigated. The majority of binary
millisecond pulsars are found to descend from systems in which the
most massive component undergoes a common-envelope phase prior to the
supernova explosion leading to the birth of the neutron star.

The orbital period distribution of simulated samples of wide binary
millisecond pulsars is compared with the observed distribution of
Galactic binary millisecond pulsars for a variety of parameters
describing the formation and evolution of binaries.  The distribution
functions typically show a short-period peak below 10 days and a
long-period peak around 100 days. The observed distribution is best
reproduced by models with highly non-conservative mass transfer,
common-envelope efficiencies equal to or larger than unity, a critical
mass ratio for the delayed dynamical instability larger than 3, and no
or moderate supernova kicks at the birth of the neutron star. Few
systems are found with orbital periods longer than 200 days,
irrespective of the accretion efficiency of neutron stars. This occurs
as a result of the upper limit on the initial orbital periods beyond
which the binary avoids the common-envelope phase prior to the
supernova explosion of the primary.
\end{abstract}


\section{Introduction}

In close binaries containing a neutron star, the evolution of the
system is driven by the nuclear evolution of the companion and the
loss of angular momentum through processes as magnetic braking and
gravitational radiation. When the combination of these processes
causes the companion to fill its Roche lobe, mass and angular momentum
are transferred to the neutron star which is then spun up to form a
recycled pulsar, unless transient behaviour prevents effective
accretion. At the end of the mass-transfer phase, the binary emerges
as a binary millisecond pulsar (BMSP) consisting of a rapidly rotating
neutron star in a nearly circular orbit around a low-mass white dwarf,
which used to be the core of the Roche-lobe overflowing companion (for
a review see Bhattacharya \& van den Heuvel 1991, Phinney \& Kulkarni
1994). If the mass transfer phase takes place during the companion's
ascent of the red giant branch, the relation between the core mass and
the radius of the star and the relation between the Roche-lobe radius
and the orbital separation, lead to a correlation between the orbital
period of the BMSP and the mass of the white dwarf. This correlation
has been studied extensively by Joss, Rappaport \& Lewis (1987),
Savonije (1987), Rappaport et al. (1995), Ritter (1999), and Tauris \&
Savonije (1999).

The precise evolutionary path followed by a BMSP progenitor depends
mainly on the orbital period and on the mass of the donor star at the
onset of Roche-lobe overflow after the supernova explosion of the
primary. Systems with orbital periods longer than $\sim 20$ days and
donors more massive than $\sim 2.0\,M_\odot$ are likely to evolve
through a common-envelope phase during which the envelope of the donor
is expelled and the neutron star spirals in towards the donor star's
core (Kalogera \& Webbink 1998). If the orbital period is shorter, the
common-envelope phase can be avoided for donor stars up to $\sim
4.0\,M_\odot$ (Tauris, van den Heuvel \& Savonije 2000; Kolb et
al. 2000; Podsiadlowski, Rappaport \& Pfahl 2002). Binaries with donor
stars less massive than $\sim 2.0\,M_\odot$ will become low-mass X-ray
binaries, for which the evolution depends on whether the orbital
period is longer or shorter than the bifurcation period separating
diverging from converging systems (Pylyser \& Savonije 1988,
1989). Only low-mass X-ray binaries with orbital periods longer than
the bifurcation period will eventually evolve into wide BMSPs.

At present, 33 BMSPs with orbital periods longer than one day have
been found in the Galactic disc. Their orbital periods,
eccentricities, and white dwarf masses are listed in Table~\ref{bmsp}
(Ritter, private communication). Inspection of the observed orbital
periods shows a paucity of systems with periods between 30 and 60
days, and an absence of systems with periods longer than 200 days. A
theoretical interpretation of the possible period gap was given by
Tauris (1996) and Taam, King \& Ritter (2000), while an explanation
for the upper limit was put forward by Ritter \& King (2002).

\begin{table}
\caption{Observed orbital periods, eccentricities, and white dwarf
  masses of Galactic BMSPs with orbital periods longer than 1
  day. The data is derived from Taam et al. (2000) with updated
  information from Ritter (private communication).}
\label{bmsp}
\begin{tabular}{lr@{.}lr@{.}lr@{.}l}
\hline
Name   & \multicolumn{2}{c}{$P_{\rm orb}$ (days)} &  \multicolumn{2}{c}{$e$} & \multicolumn{2}{c}{$M_{\rm WD}/M_\odot$} \\
\hline
J0613-0200  &    1&19851 &  0&000007 &  $>$0&13  \\
J1435-6100  &    1&35489 &  0&000010 &  $>$0&90  \\
J0034-0534  &    1&58928 & $<$0&00002  & $>$0&15  \\
J0218+4232  &    2&02885 & $<$0&00002  & $>$0&16  \\
J2317+1439  &    2&45933 & 0&0000005& \multicolumn{2}{c}{} \\
J1911-1114  &    2&71656 & $<$0&000013 & $>$0&12 \\
J1157-5112  &    3&50739 &  0&000402 &  $>$1&18  \\
J1045-4509  &    4&08353 &  0&000024 &  $>$0&16  \\
J1745-0952  &    4&94345 &  0&000018 & $>$0&11   \\
J1732-5049  &    5&26300 &  0&000001 &  0&18       \\
J0437-4715  &    5&74104 &  0&000019 &  \multicolumn{2}{c}{0.22-0.32} \\
J1603-7202  &    6&30863 & $<$0&00002  & $>$0&29  \\
J2129-5721  &    6&62549 & $<$0&000017 & $>$0&14 \\
J2145-0750  &    6&83890 & 0&000018 & $>$0&43 \\
J1022+1001  &    7&80513 &  0&000098 &  $>$0&73  \\
J0621+1002  &    8&31868 &  0&002458 &  $>$0&45  \\
J1518+4904  &    8&63400 &  0&249485 &  \multicolumn{2}{c}{}  \\
J1918-0642  &   10&91318 &  0&000022 & $>$0&24   \\
J1804-2717  &   11&12871 &  0&000035 & $>$0&21   \\
B1855+09    &   12&32717 &  0&000027 &  \multicolumn{2}{c}{0.24-0.29} \\
J1454-5846  &   12&42307 &  0&001898 &  $>$0&87  \\
J1810-2005  &   15&01220 &  0&000025 & $>$0&28   \\
J1709+23    &   22&7    & \multicolumn{2}{c}{} & \multicolumn{2}{c}{} \\
J1618-3919  &   22&8     & \multicolumn{2}{c}{} & \multicolumn{2}{c}{} \\
J2033+17    &   56&2     & $<$0&05     & $>$0&2  \\
J1713+0747  &   67&82513 &  0&000075 & $>$0&27   \\
J1455-3330  &   76&17458 &  0&000167 &  $>$0&27  \\
J2019+2425  &   76&51163 &  0&000111 &  0&33       \\
J2229+2643  &   93&01589 & 0&000256 & \multicolumn{2}{c}{} \\
J1643-1224  &  147&01740 &  0&000506 & $>$0&13   \\
B1953+29    &  117&34910 &  0&000330 & \multicolumn{2}{c}{}  \\
J1640+2224  &  175&46066 &  0&000797 & \multicolumn{2}{c}{} \\
\hline
\end{tabular}
\end{table}

With this paper, we initiate a systematic study based on population
synthesis techniques to investigate the different evolutionary
channels leading to the formation of wide BMSPs and to compare the
theoretical with the observed orbital period distribution of wide
BMSPs. We particularly investigate whether a set of standard
evolutionary parameters can be found to reproduce the observed orbital
period distribution without including observational selection effects
or pulsar lifetime issues. Due to the large uncertainties involved in
the determination of the masses of the white dwarfs in the observed
BMSPs (see Table~\ref{bmsp}) we presently do not consider the white
dwarf mass distribution as good a tool as the orbital period
distribution to confine stellar and binary evolution and formation
parameters.

Since BMSPs located in globular clusters are thought to be formed
through tidal encounters of neutron stars with primordial binaries
(e.g. Rappaport, Putney \& Verbunt 1989), we limit ourselves to BMSPs
in the Galactic disc. We also leave aside systems with orbital periods
shorter than 1 day since they are likely to follow evolutionary paths
different from those of the BMSPs investigated here. In particular,
the evolution of the short-period systems is dominated by angular
momentum losses for which quantitative descriptions are still
uncertain (Ergma \& Sarna 1996; Kalogera, Kolb \& King 1998;
Ergma, Sarna \& Antipova 1998; Rasio, Pfahl \& Rappaport 2000; Dewi et
al. 2002; and in particular Podsiadlowski et al. 2002 and 
references therein).

The plan of the paper is as follows. In Sect.~2, we briefly summarise
the basic concepts of the binary evolution code used in this
investigation.  In Sect.~3, the evolutionary channels leading to the
formation of wide BMSPs are discussed. In Sect.~4, we explore the
parameter space occupied by the BMSP progenitors during various phases
of their evolution. The effects of the input parameters adopted in our
binary evolution calculations on the orbital period distribution of
wide BMSPs and the relative contributions of the different formation
channels to the total population of wide BMSPs are investigated in
Sect.~5. The final section is devoted to concluding remarks.

\section{The binary evolution code}

\subsection{Single star evolution}

We developed a rapid binary evolution code based on the analytical
approximations for the evolution of single stars with masses
ranging from $0.1$ to $100\,M_\odot$ and metallicities between $Z =
0.0001$ and $Z = 0.03$ derived by Hurley, Pols \& Tout (2000). The
formulae express the luminosity, radius, and core mass of a
star as a function of its age, mass, and metallicity for all phases
from the zero-age main sequence up to and including the remnant
stages.  For massive stars ending their life as a neutron star, we
simplified the prescription for the mass of the remnant so that all
neutron stars are born with a mass of $1.4\,M_\odot$. The
simplification is justified by the fact that most neutron stars with
accurately known masses have a mass close to this value (Thorsett \&
Chakrabarty 1999). We furthermore limit ourselves to Population~I
stellar compositions.

The single star evolution scheme takes into account mass loss from the
envelope due to stellar winds. Stars beyond the main sequence are
subjected to the mass-loss rates given by Kudritzki \& Reimers (1978),
the superwind phase on the asymptotic giant branch (AGB) is modelled
using the prescription of Vassiliadis \& Wood (1993), massive
stars are subjected to mass loss over the entire Hertzsprung-Russell
diagram as outlined by Nieuwenhuizen \& De Jager (1990) and Kudritzki
et al. (1989), and naked helium stars are subjected to a
Wolf-Rayet-type mass-loss rate. For more details, we refer to Hurley
et al.\ (2000).

In addition to the stellar evolution of the binary components, we also
take into account the evolution of the binary configuration due to
gravitational radiation, magnetic braking, mass loss and accretion
from stellar winds, and conservative or non-conservative Roche-lobe
overflow. For the implementation of magnetic braking, we adopt a 
Skumanich-type parametrisation modified to take into account the
dependency on the mass of the convective envelope as described by
Hurley et al. (2000). We furthermore assume the orbit of
the binary to be circular and keep the rotation of the stars
synchronised with their orbital motion at all times. The recipe
followed for the implementation of the binary evolution is essentially
along the lines laid down by Hurley, Tout \& Pols (2002). In the
following paragraphs we therefore restrict ourselves to summarising
the points where our treatment deviates from the outline given by
these authors.

\subsection{Roche-lobe overflow}

When the combined effects of stellar and orbital evolution cause a
component of a close binary to become larger than its Roche lobe, mass
is transferred from the Roche-lobe overflowing star to the
companion. The stability of the mass transfer process depends on the
adiabatic and thermal radius-mass exponents, $\zeta_{\rm ad}$ and
$\zeta_{\rm th}$, and the Roche-lobe index $\zeta_{\rm L}$ (for a
definition see, e.g., Webbink 1985). For stars on the main sequence or
in the Hertzsprung gap, we use the tabulated values for $\zeta_{\rm
ad}$ and $\zeta_{\rm th}$ given by Hjellming (1989). In all other
cases we adopt the prescription given by Hurley et al.\ (2002). For
the determination of the Roche-lobe index $\zeta_{\rm L}$, we follow
the procedure outlined by Kolb et al. (2001).

If mass transfer proceeds on either the nuclear ($\zeta_{\rm L} <
\zeta_{\rm ad}$ and $\zeta_{\rm L} < \zeta_{\rm eq}$) or the thermal
($\zeta_{\rm eq} < \zeta_{\rm L} < \zeta_{\rm ad}$) timescale of the
Roche-lobe overflowing star, the companion is assumed to accrete a
fraction $1-\gamma$ of the transferred mass determined by the ratio of
the mass-transfer timescale to the accretor's thermal timescale (for
details see Hurley et al.\ 2002). The remaining fraction is assumed to
be expelled from the system, carrying away the specific orbital
angular momentum of the companion. In the particular case of a neutron
star accretor, an average accretion rate is determined by imposing an
upper limit of $\left(\Delta M_{\rm NS} \right)_{\rm max}$ on the
amount of mass that can be accreted (for details see
Appendix~\ref{rlof}). Once this limit has been reached, mass accretion
onto the neutron star is switched off completely to ensure that the
neutron star accretes no more than $\left(\Delta M_{\rm NS}
\right)_{\rm max}$. We impose this upper limit to mimic the low
accretion efficiency neutron stars are thought to have (e.g. King \&
Ritter 1999, Ritter \& King 2002). In addition, we also limit the
mass-accretion rate to the Eddington rate at all times.

In the case of dynamically unstable mass transfer ($\zeta_{\rm ad} <
\zeta_{\rm L}$) from a giant-like star, the binary is assumed to enter
a common-envelope phase during which the orbital separation is reduced
and the envelope is ejected from the system. The phase is modelled by
equating the binding energy of the giant's envelope to the change in
orbital energy of the two stars:
\begin{equation}
{{G \left( M_{\rm c} + M_{\rm e} \right) M_{\rm e}}
  \over {\lambda\, R_{\rm L}}} =
  \alpha \left[ {{G\, M_{\rm c}\, M_2} \over {2\, a_{\rm f}}}
  - {{G \left( M_{\rm c} + M_{\rm e} \right) M_2}
  \over {2\, a_{\rm i}}} \right]\!.  \label{ce}
\end{equation}
In this equation, $G$ is the gravitational constant, $M_{\rm c}$ and
$M_{\rm e}$ are the core and envelope mass of the Roche-lobe
overflowing star, $R_{\rm L}$ is the radius of the donor star's Roche
lobe, $M_2$ is the mass of the companion, $a_{\rm i}$ and $a_{\rm f}$
are the orbital separations of the binary at the start and at the end
of the common-envelope phase, $\lambda$ is a dimensionless
structure parameter determining the binding energy of the giant's
envelope, and $\alpha$ is the fraction of the orbital energy that is
transferred to the envelope. The parameters $\lambda$ and
$\alpha$ are usually assumed to be constant and equal to 0.5 and 1.0,
respectively (e.g.\ de Kool 1992, Politano 1996). In fact, the
value of $\lambda$ depends on the evolutionary stage of the donor star
at the onset of the mass-transfer phase (Dewi \& Tauris 2000,
2001). However, since no elaborate set of $\lambda$-values has been
published for the full range of stellar masses considered in our 
investigation, we here treat $\lambda$ as a constant.

\subsection{Supernova explosions}

Whenever a component of a close binary undergoes a supernova
explosion leading to the birth of a neutron star, we use the
expressions established by Kalogera (1996) to derive the
post-supernova orbital parameters. If the birth of the neutron star is
accompanied by a kick, we assume its magnitude $v_{\rm kick}$ to be
distributed according to the Maxwellian distribution
\begin{equation}
P \left( v_{\rm kick} \right) = \sqrt{2 \over \pi}\,
  {v_{\rm kick}^2 \over {\sigma^3}}\, \exp \left(
  - {v_{\rm kick}^2 \over {2\, \sigma^2}} \right), \label{Pkick}
\end{equation}
where $\sigma$ is the dispersion of the kick velocities. If the
supernova explosion causes the binary to become eccentric, we
instantaneously circularise the orbit while conserving the total
orbital angular momentum of the binary.

\section{Evolutionary channels}

Before proceeding to the statistics of the BMSP population and the
effects of the various assumptions made in our treatment of stellar
and binary evolution and formation, we explore the different
evolutionary channels leading to the formation of wide BMSPs.

We followed the evolution of a large number of binaries starting with
zero-age main sequence components up to a maximum evolutionary age of
15\,Gyr. The initial masses of the binary components are taken in the
interval between $0.1\,M_\odot$ and $60\,M_\odot$, and the initial
orbital periods in the range from 10 to 10\,000 days. We used
equidistantly spaced grids consisting of 40 grid points for the
logarithm of the initial masses and 200 grid points for the logarithm
of the initial periods. For symmetry reasons only binaries with $M_1 >
M_2$ are evolved. Each time a supernova explosion leads to the birth
of a neutron star, 200 random kick velocities are generated and the
evolution of the surviving binaries is followed until the imposed age
limit of 15\,Gyr. In order to identify as many formation channels as
possible, we do not apply any weighting to the initial masses and the
initial orbital periods at this stage. The weighted contributions of
the different formation channels to the population of wide BMSPs will
be discussed in Sect.~5.

For the results presented in this section, a standard set of input
parameters was adopted for the binary evolution calculations:
common-envelope phases are modelled using the values $\lambda=0.5$ and
$\alpha=1.0$ for the binding-energy parameter and the
envelope-ejection efficiency, the occurrence of a delayed dynamical
instability is determined using the values tabulated by Hjellming
(1989), supernova kicks are generated with a Maxwellian velocity
dispersion $\sigma=190\, {\rm km/s}$, and an upper limit $\left(\Delta
M_{\rm NS} \right)_{\rm max} = 0.2\,M_\odot$ is imposed on the mass
that can be accreted by a neutron star. These assumptions will serve
as a reference for the population synthesis calculations presented in
Sect.~5, where we will refer to them as model~A.

We find four major evolutionary channels leading to the formation of
wide BMSPs. In three of these channels, the primary undergoes a 
common-envelope phase (CE) which ejects the primary's envelope from the
system and reduces the orbital separation. For primaries more massive
than $\sim\! 20\, M_\odot$ the common-envelope phase takes place when
the primary crosses the Hertzsprung gap, while for primaries less
massive than $\sim\! 20\, M_\odot$ the common-envelope phase takes
place when the primary ascends the AGB. Both cases lead to similar
evolutionary paths. For the continuation of the evolution, it is
necessary to distinguish between binaries with low- and intermediate
mass secondaries.

\begin{figure}
\resizebox{8.0cm}{!}{\includegraphics{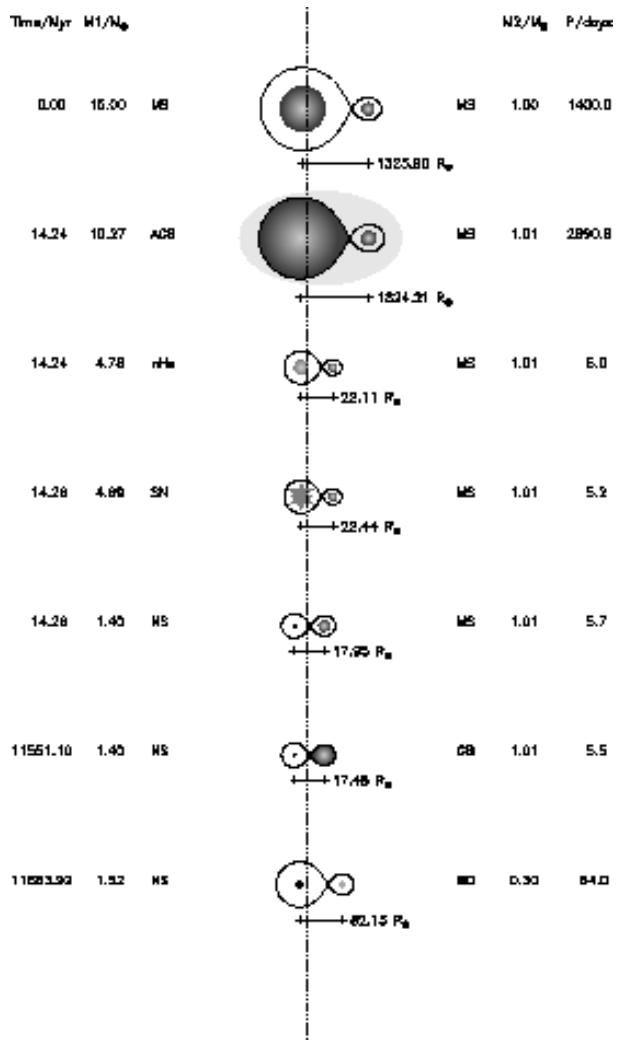}}
\caption{The main evolutionary phases of a BMSP progenitor with a
  low-mass secondary evolving into a helium white dwarf [channel
  He~N]. The two distinguishing phases for this formation channel are
  the CE phase of the primary prior to the supernova explosion and the
  stable case~B mass-transfer phase after the explosion. The dotted
  vertical line indicates the mass centre of the binary. }
\label{He23}
\end{figure}

The main evolutionary phases of a BMSP progenitor with a low-mass
secondary are illustrated in Fig.~\ref{He23}, where the initial orbital
period is 1400 days and the binary components have initial masses of
$15\,M_\odot$ and $1\,M_\odot$. The primary fills its Roche lobe after
14\,Myr, at which time a stellar wind has reduced the mass of the
primary to $10.3\,M_\odot$ and increased the orbital period to $\sim
2690$ days. The primary exits the common-envelope phase as a massive
naked helium star which evolves further until it explodes in a type
Ib/c supernova and gives birth to a $1.4\,M_\odot$ neutron star. After
circularisation of the post-supernova orbit, the binary consists of a
$1.0\,M_\odot$ main-sequence star which orbits the neutron star in a
period of 5.7 days. The binary subsequently evolves as a detached
system until the secondary expands and fills its Roche lobe during its
ascent of the giant branch. The resulting mass-transfer phase is both
dynamically and thermally stable and takes place on the nuclear
timescale of the giant star. During this phase, mass accretion spins
up the neutron star to high rotation rates until it becomes a recycled
pulsar. At the end of the mass-transfer phase, the binary emerges as a
wide BMSP consisting of a rapidly rotating neutron star in a 64 day
orbit around a $0.3\,M_\odot$ helium white dwarf, the former core of
the giant star. This formation channel has been discussed previously
by Bhattacharya \& van den Heuvel (1991) and Phinney \& Kulkarni
(1994), among others. We will refer to it with the label He~N.

\begin{figure}
\resizebox{8.0cm}{!}{\includegraphics{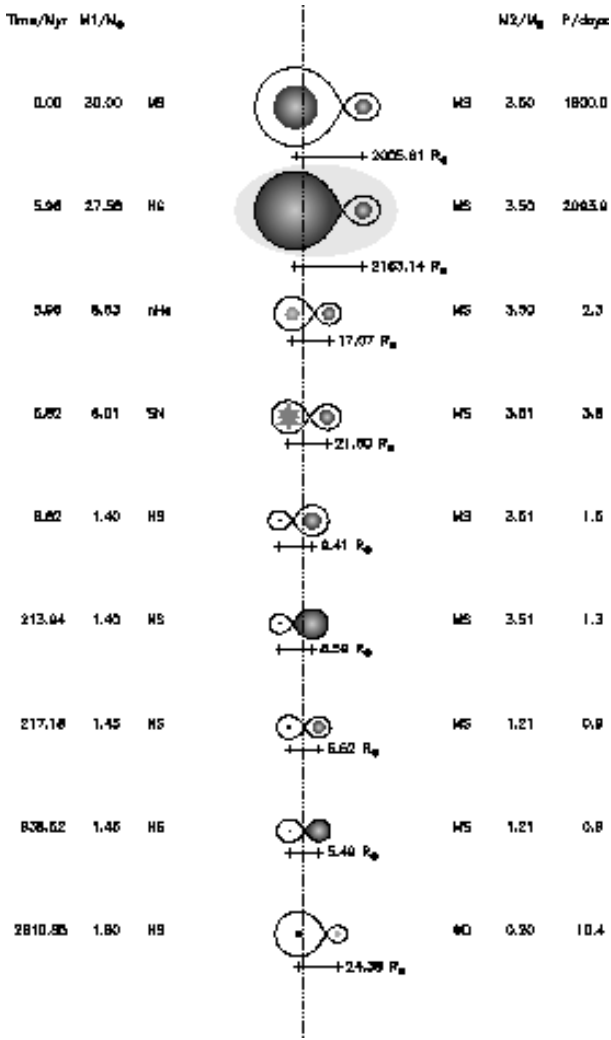}}
\caption{As Fig.~\ref{He23}, but with an intermediate-mass secondary
  evolving into a helium white dwarf [channel He~T]. The
  distinguishing phases are the CE phase of the primary prior to the
  supernova explosion and the initial thermal-timescale case~A
  mass-transfer phase after the explosion. }
\label{He22}
\end{figure}

\begin{figure}
\resizebox{8.0cm}{!}{\includegraphics{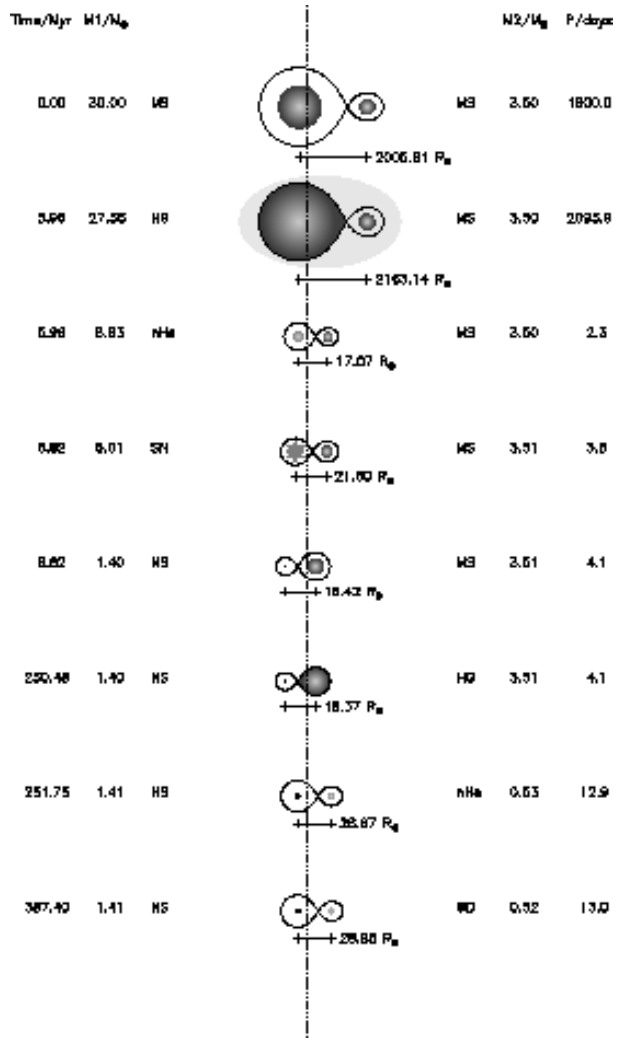}}
\caption{Same as Fig.~\ref{He22}, but with a kick velocity such that
  the secondary evolves into a carbon/oxygen white dwarf [channel
  CO~T]. The distinguishing phases are the CE phase of the primary
  prior to the supernova explosion and the thermal-timescale
  case~B mass-transfer phase after the explosion. }
\label{CO22}
\end{figure}

For BMSP progenitors with intermediate-mass secondaries, the formation
channel becomes a bit more complex. Two possible evolutionary paths
are displayed in Figs.~\ref{He22} and~\ref{CO22}, in the case of a
binary with an
initial orbital period of 1400 days and initial component masses of
$30\,M_\odot$ and $3.5\,M_\odot$. Again, the primary emerges from the
common-envelope phase as a massive naked helium star which gives birth
to a $1.4\,M_\odot$ neutron star in a type Ib/c supernova
explosion. The further evolution depends on the circularised
post-supernova orbital separation.

If the post-supernova orbital period is smaller than $\sim 2.5$ days
(Fig.~\ref{He22}), the binary enters a case~A thermal-timescale
mass-transfer phase after 214\,Myr and detaches again after 217\,Myr
when the donor mass has decreased to $1.2\,M_\odot$. This short
thermal-timescale mass-transfer phase is followed by a stable nuclear
timescale mass-transfer phase after 939\,Myr when the secondary
approaches the end of the main sequence. This stable mass transfer
phase continues all the way up to the giant branch and leads to the
formation of a binary consisting of a rapidly rotating neutron star in
a 10 day orbit around a $0.2\,M_\odot$ helium white dwarf. We label
this formation channel He~T.

If the post-supernova orbital period is larger than $\sim 2.5$ days
(Fig.~\ref{CO22}), the secondary enters a case~B 
thermal-timescale mass-transfer phase when it crosses the Hertzsprung
gap. Due to the high mass-transfer rate, the neutron star rejects most
of the mass transferred by the companion so that the spin-up process
is probably less effective and thus leads to more slowly rotating
pulsars than the spin-up process during a slow stable mass-transfer
phase (e.g. Li 2002, and references therein). The semi-detached state
ends when the hydrogen envelope of the secondary is completely
stripped away and its helium core is exposed as a low-mass naked
helium star. During the following 135\,Myr, the secondary burns helium
in the core while a Wolf-Rayet-type stellar wind slowly removes the
outer helium layers. The system ends up as a $0.5\,M_\odot$
carbon/oxygen white dwarf in a 10.4 day orbit around a neutron star
which may or may not be spun up to millisecond periods. This
channel was first found independently by King \& Ritter (1999) and by
Podsiadlowski \& Rappaport (2000) [see also Tauris et al. (2000)]. A 
possible BMSP progenitor that is currently at the end of the early
case~B mass transfer phase in this channel is the persistent X-ray
binary Cygnus X-2 (King \& Ritter 1999, Kolb et al. 2000,
Podsiadlowski \& Rappaport 2000). We will refer to this formation
channel with the label CO~T.

\begin{figure}
\resizebox{8.0cm}{!}{\includegraphics{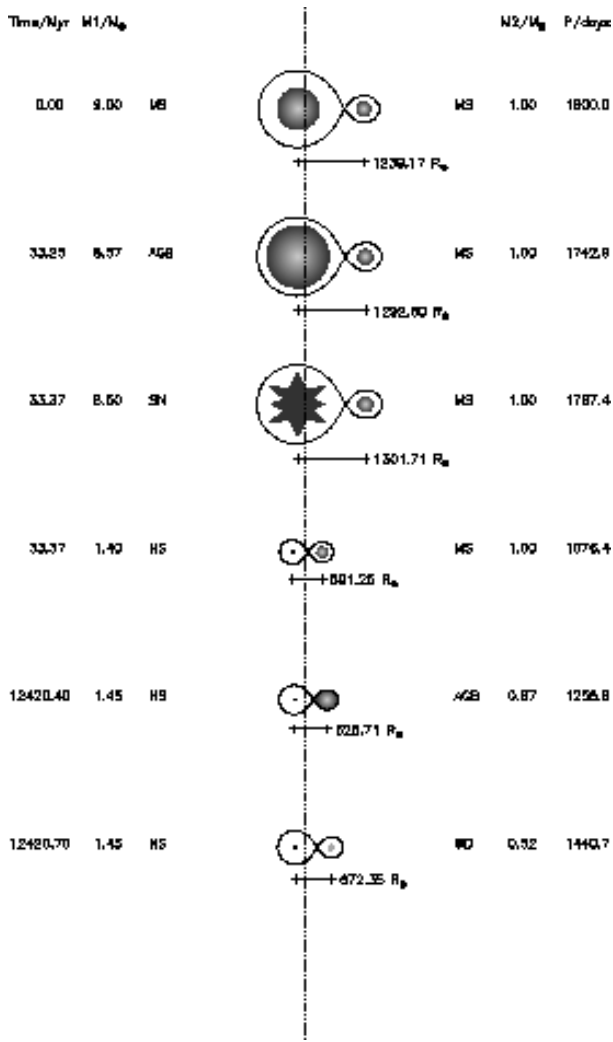}}
\caption{As Fig.~\ref{He23}, but with a low-mass secondary evolving
  into a carbon/oxygen white dwarf [channel CO~N]. The distinguishing
  characteristic is the supernova prior to any interaction
  between the binary components (see also Kalogera 1998). }
\label{CO23}
\end{figure}

The fourth and final evolutionary channel for the formation of wide
BMSPs corresponds to the direct supernova mechanism discussed by
Kalogera (1998). It applies to binaries with low-mass secondaries and
initial orbital separations that are wide enough to avoid any kind of
Roche-lobe overflow before the supernova explosion of the primary. An
example of such a system is displayed in Fig.~\ref{CO23}, where the
initial orbital period of the binary is 1600 days and the initial
masses of the primary and the secondary are $9\,M_\odot$ and
$1\,M_\odot$, respectively. The wide orbit allows the primary to
evolve all the way up to the AGB where it explodes in a type II
supernova. After the circularisation of the post-supernova orbit, the
binary consists of a $1.4\,M_\odot$ neutron star orbiting a fairly
unevolved main-sequence star with a long orbital period of 1100
days. The secondary then in turn evolves up to the AGB where it fills
its Roche lobe after 12.4\,Gyr. At this point, its mass is reduced to
$0.7\,M_\odot$ due to strong stellar winds on the giant branch and the
AGB. The following mass-transfer phase is
dynamically and thermally stable, and lasts for about 0.3\,Myr.  At
the end of the mass-transfer phase, the binary consists of a rapidly
rotating neutron star in a 1440 day orbit around a $0.5\,M_\odot$
carbon/oxygen white dwarf. We label this formation channel CO~N.

\section{Parameter space}

Next, we examine the parameter space occupied by systems forming wide
BMSPs at the start of their evolution, at the instant just before the
supernova explosion of the primary, at the onset of Roche-lobe
overflow after the supernova explosion, and at the birth of the BMSP.

\begin{figure*}
\resizebox{8.5cm}{!}{\includegraphics{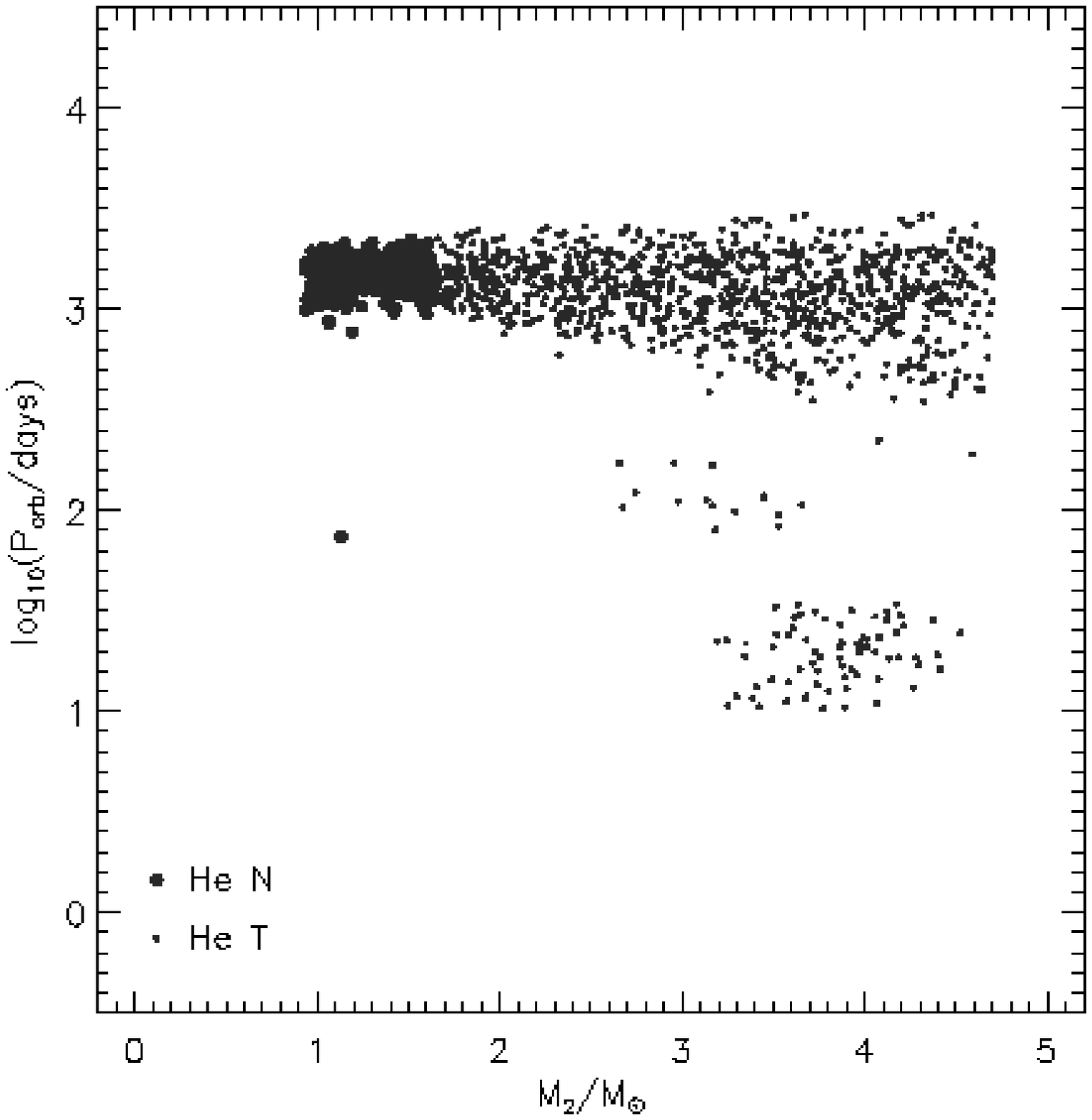}}
\resizebox{8.5cm}{!}{\includegraphics{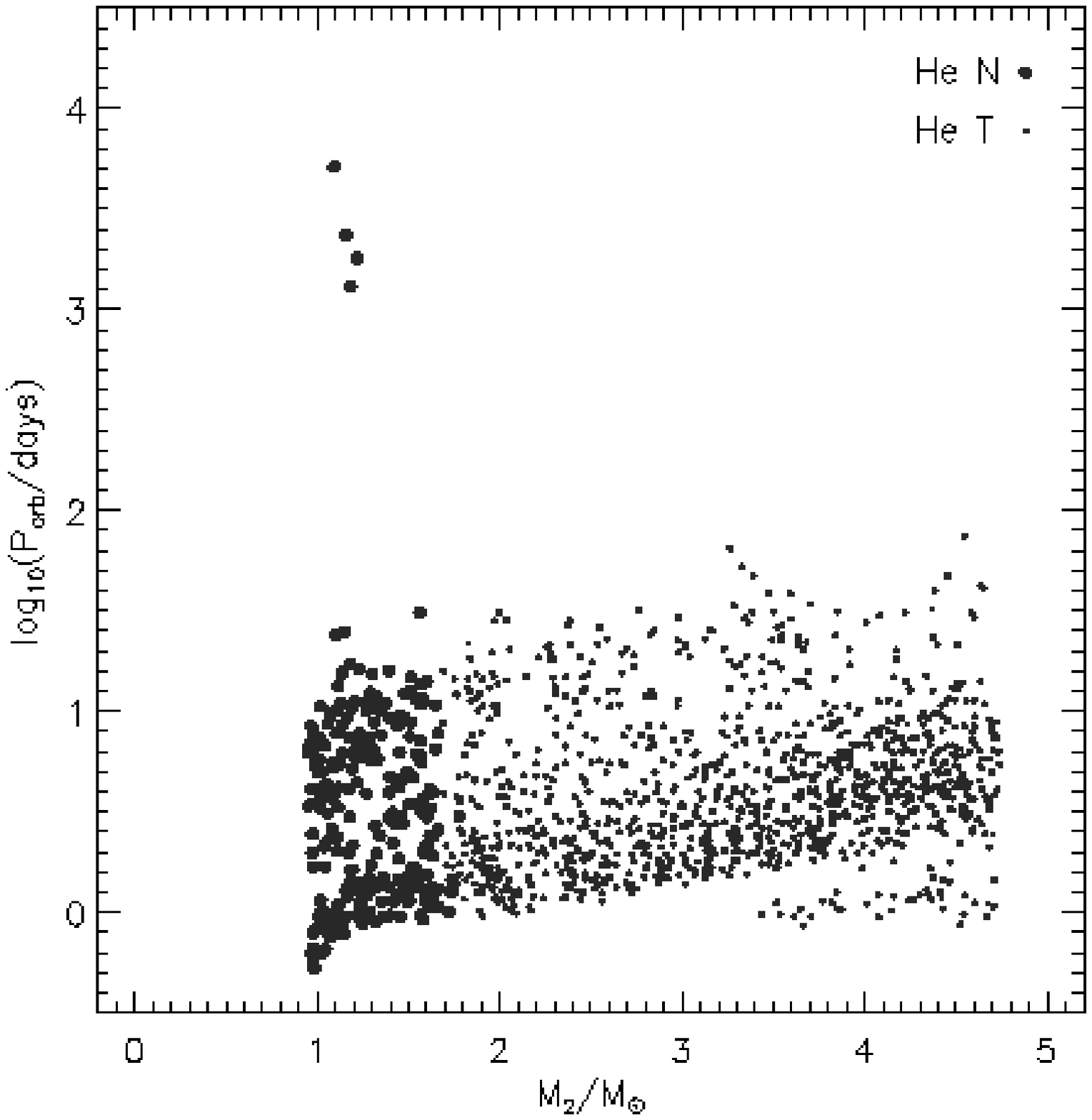}}
\resizebox{8.5cm}{!}{\includegraphics{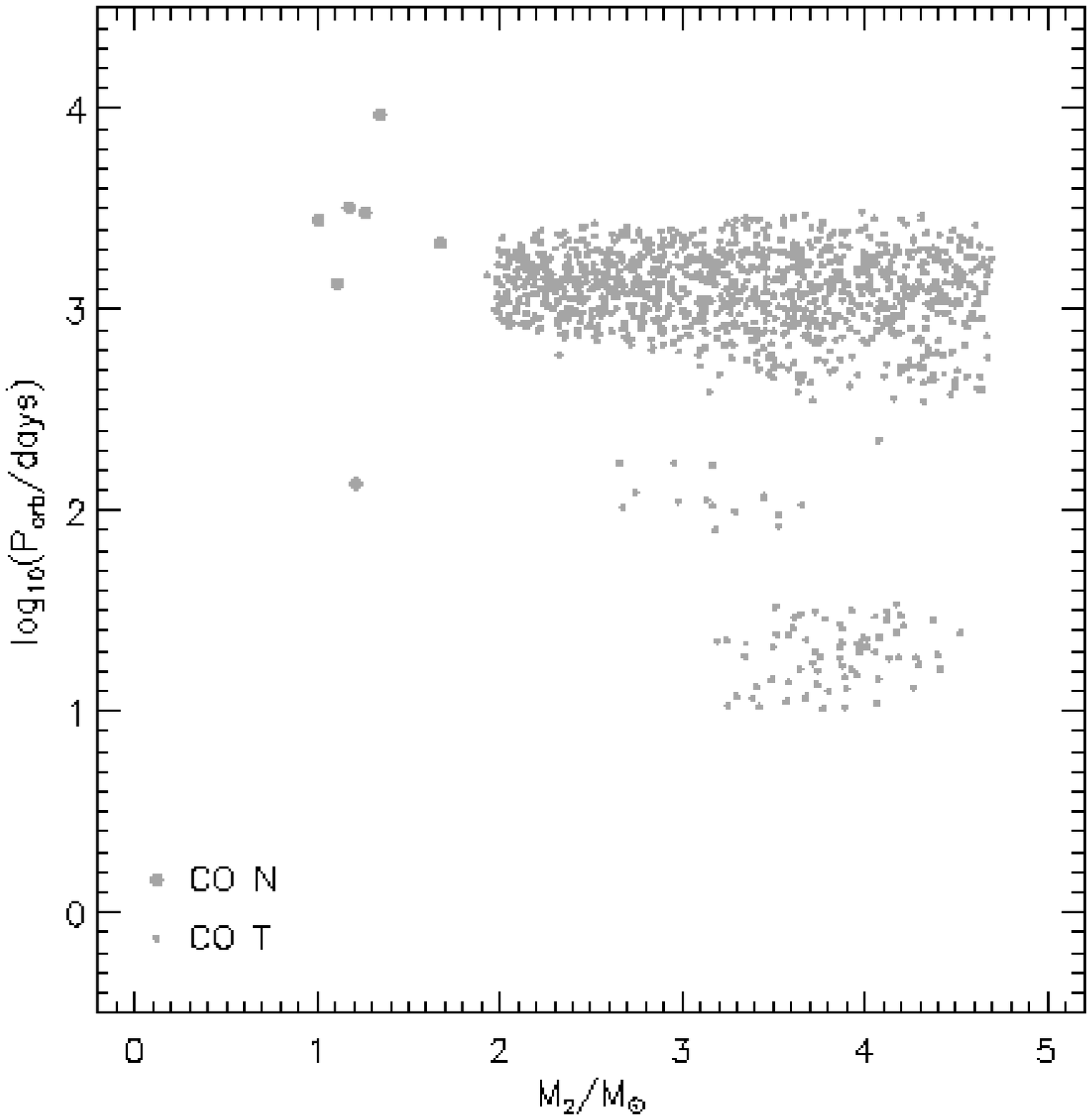}}
\resizebox{8.5cm}{!}{\includegraphics{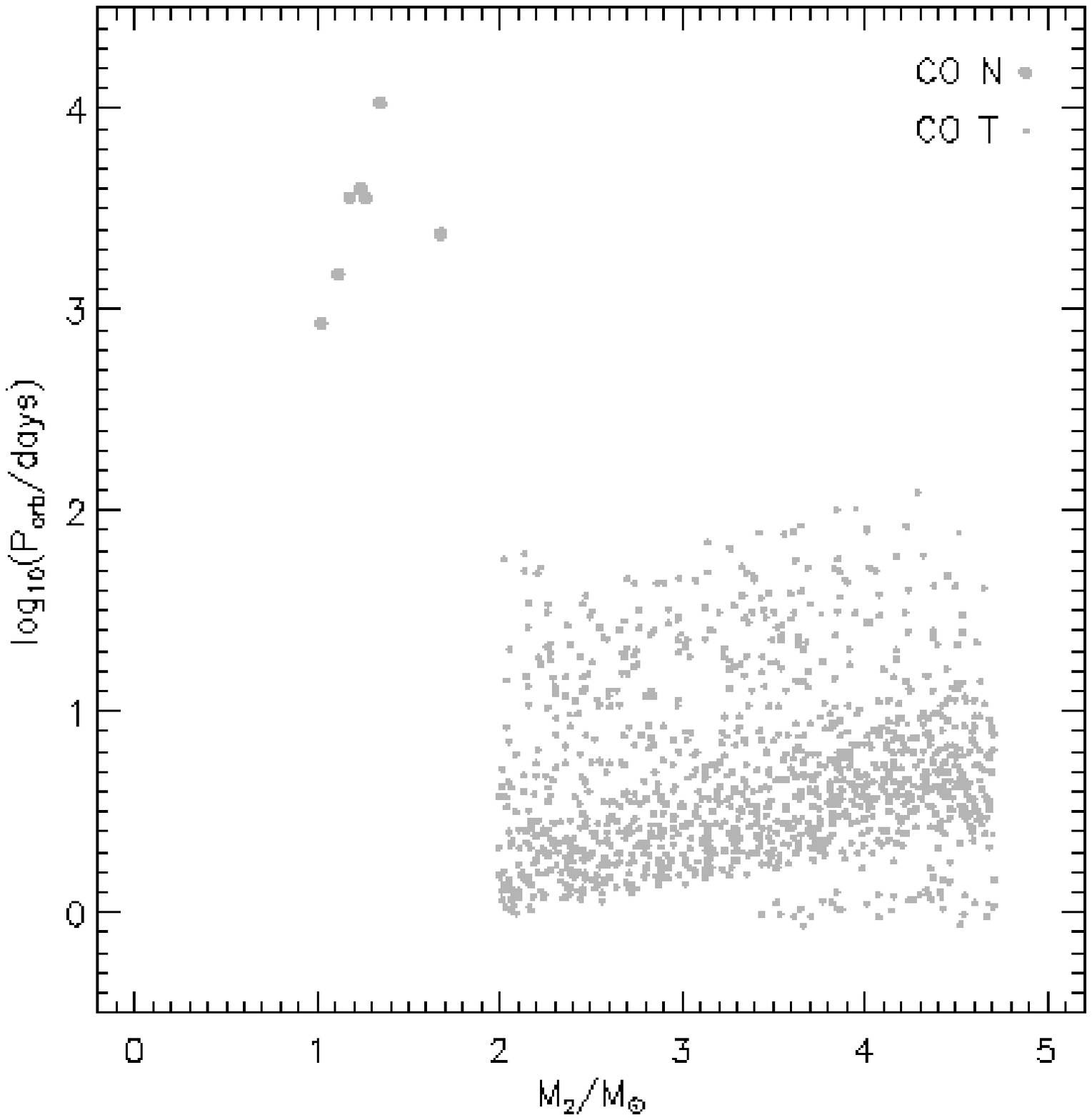}}
\caption{The initial (left) and pre-supernova (right) parameter space
  occupied by the BMSP progenitors. Small dots represent systems
  undergoing a thermal-timescale mass-transfer phase after the
  supernova explosion of the primary, while big dots represent systems
  for which mass transfer from the secondary is stable at all
  times. Dark and light dots correspond to systems leading to BMSPs
  containing helium and carbon/oxygen white dwarfs, respectively. For
  clarity, the BMSPs containing helium and carbon/oxygen white dwarfs
  are shown in separate panels.}
\label{presn}
\end{figure*}

The initial orbital periods $P_{\rm orb}$ and secondary masses $M_2$
of binaries that will evolve into BMSPs are shown in the left-hand
panels of Fig.~\ref{presn}. Dark and light dots indicate
systems evolving into BMSPs containing a helium white dwarf and
systems evolving into BMSPs containing a carbon/oxygen white dwarf,
respectively, 
while big and small dots distinguish between systems undergoing a
nuclear or thermal-timescale mass-transfer phase after the supernova
explosion of the primary. The bulk of the BMSP progenitors occupy a
rather narrow band of orbital periods around 1000 days. The upper
limit of the band corresponds to the longest orbital periods for which
the primary fills its Roche lobe and undergoes a common-envelope phase
prior to its supernova explosion, while the lower limit is formed by the
shortest orbital periods for which mergers or contact configurations
can be avoided. Besides the band around 1000 days, there is also a
small group of systems with orbital periods in the range from 10 to 30
days. These systems evolve along the same lines as illustrated in
Figs.~\ref{He22} (the He T channel) and~\ref{CO22} (the CO T channel),
except that the primary undergoes a rapid thermal-timescale
mass-transfer phase instead of a common-envelope phase prior to its
supernova explosion. Since these systems do not contribute
significantly to the total population of BMSPs we do not consider them
separately in our present investigation.

The right-hand panels of Fig.~\ref{presn} show the orbital periods and
secondary masses of the BMSP progenitors just before the supernova
explosion of the primary. With the exception of a few rare cases, the
orbital separations of the initial binaries are substantially reduced
due to the common-envelope phase of the primary. The few systems
managing to avoid the common-envelope phase are those evolving through
the direct supernova mechanism (Fig.~\ref{CO23}, the CO~N channel).
Their orbital periods have increased slightly with respect to their
initial orbital periods due to mass loss from the system caused by the
stellar wind of the primary. The systems evolving through the direct
supernova mechanism into BMSPs containing a helium white dwarf are
included in the He~N channel, since their post-supernova evolutionary
path is similar to that shown in Fig.~\ref{He23}.

\begin{figure}
\resizebox{\hsize}{!}{\includegraphics{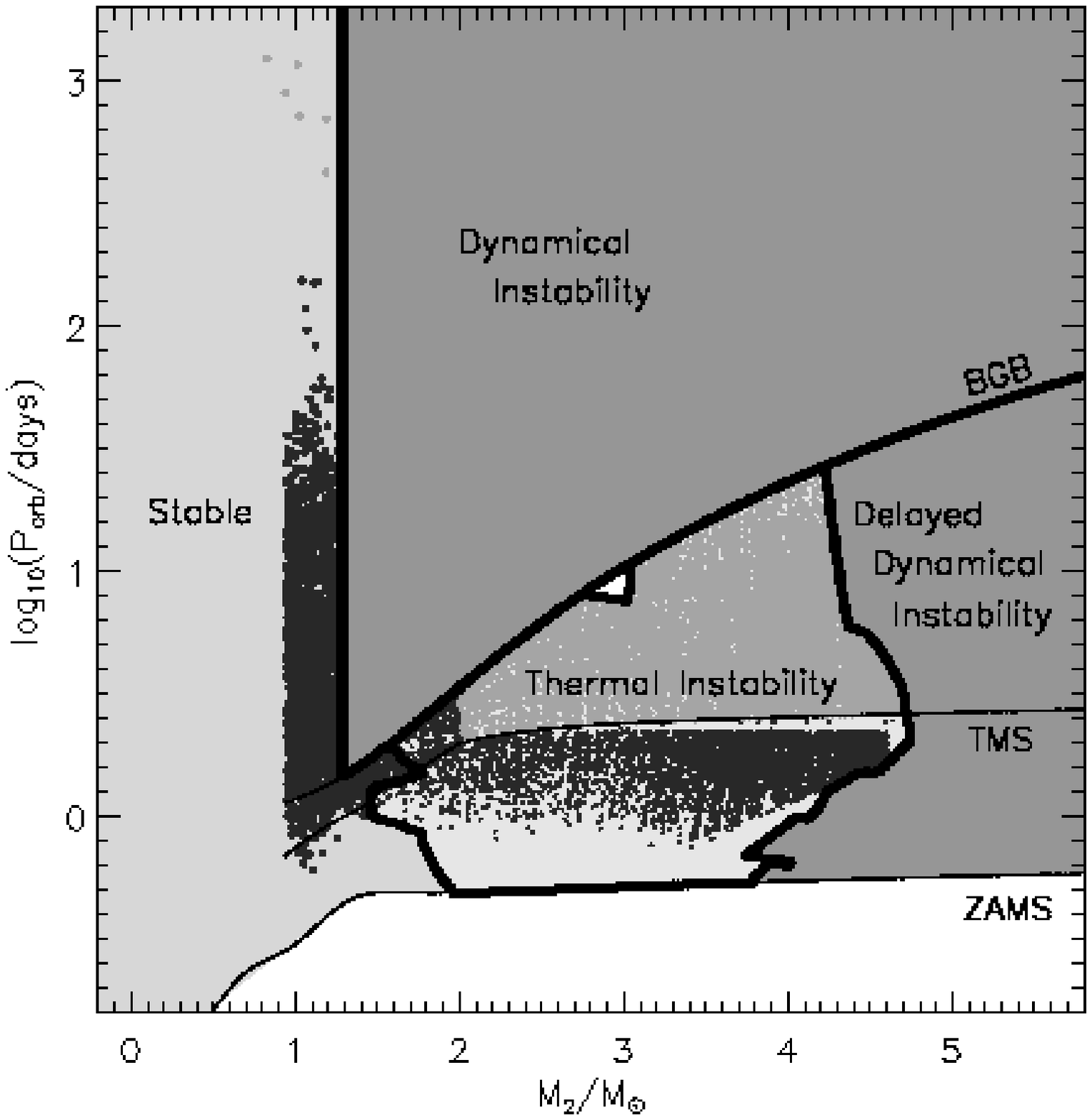}}
\caption{The parameter space occupied by the BMSP progenitors at the
  onset of Roche-lobe overflow from the secondary, after the supernova
  explosion of the primary. The different dots have the same meaning
  as in Fig.~\ref{presn}. The thin lines correspond to the orbital
  periods of a Roche-lobe filling secondary on the zero-age main-sequence
  (ZAMS), the terminal main-sequence (TMS), and the base of the giant
  branch (BGB). The thick lines delimit the different stability
  regions for mass transfer from the secondary. Dark grey indicates
  dynamically unstable or delayed dynamically unstable mass transfer,
  light grey indicates thermal-timescale mass transfer, and
  intermediate grey indicates thermally and dynamically stable mass
  transfer. For the construction of the thick and the thin lines, the
  mass of the neutron star was assumed to be $1.4\,M_\odot$. }
\label{postsn}
\end{figure}

In Fig.~\ref{postsn}, we show the parameter space occupied by the
progenitors of the wide BMSPs after the supernova explosion of the
primary, at the onset of Roche-lobe overflow from the secondary. 
The lines labelled ZAMS, TMS, and BGB represent the orbital periods
at which the secondary would fill its Roche lobe if it was located at
the zero-age main sequence, the terminal main sequence, or the base of
the giant branch, respectively. The thick lines delimit the different
stability regions for mass transfer from the secondary. The dark grey
region indicates dynamically unstable or delayed dynamically unstable
mass transfer, the light grey region indicates thermal-timescale mass
transfer, and the intermediate grey region indicates thermally and
dynamically stable mass transfer. For the construction of both thin
and thick lines, the mass of the neutron star was assumed to be
$1.4\,M_\odot$.

The post-supernova orbital periods of the BMSP progenitors depend on
the magnitude of the kick velocity imparted to the neutron star 
at birth. For secondaries with a mass larger than $\sim\! 1.3\,
M_\odot$ the dynamical instability of mass transfer sets an upper
limit for the post-supernova orbital period. The lower limit on the
orbital periods observed in Fig.~\ref{postsn} is a consequence of
the lower limit on the post-common-envelope orbital separations
resulting from the requirement that the systems exit the 
common-envelope phase as detached binaries.  Lower and upper limits on the
mass of the secondary are caused by the imposed age limit of 15\,Gyr
and by the delayed dynamical instability. Secondaries with a mass
smaller than $\sim\!0.9\, M_\odot$ evolve too slowly to fill their
Roche lobe within 15\,Gyr, while for secondaries with a mass larger
than $\sim 4.0-4.5\,M_\odot$ the initial thermal-timescale
mass-transfer phase will evolve into a dynamically unstable phase
shortly after the onset of Roche-lobe overflow.

Furthermore, four groups of systems can be distinguished in
Fig.~\ref{postsn}, in accordance with the four identified formation
channels. Systems evolving into BMSPs containing a helium white dwarf
either have low-mass secondaries which enter a stable mass-transfer
phase when they cross the Hertzsprung gap or ascend the giant branch
(Fig.~\ref{He23}, the He~N channel), or intermediate-mass secondaries
for which mass transfer initially takes place on a thermal timescale
(Fig.~\ref{He22}, the He~T channel). Systems evolving into BMSPs
containing a carbon/oxygen white dwarf undergo a 
thermal-timescale mass-transfer phase when the secondary crosses the
Hertzsprung gap (Fig.~\ref{CO22}, the CO~T channel) or a slow stable
mass-transfer phase on the AGB if they manage to avoid the
common-envelope phase of the primary prior to the supernova explosion
(Fig.~\ref{CO23}, the CO~N channel).

\begin{figure}
\resizebox{\hsize}{!}{\includegraphics{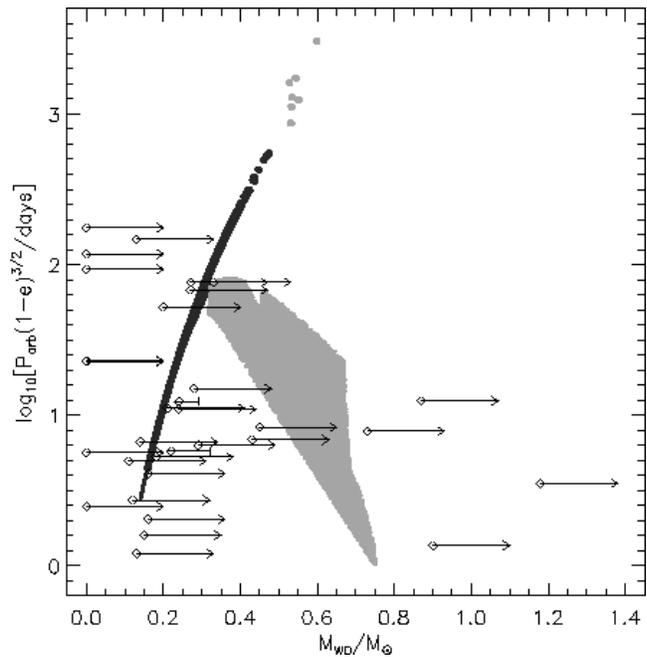}}
\caption{The parameter space occupied by wide BMSPs in the $\log
  P_{\rm orb}$ - $M_{\rm WD}$ plane.  The diamonds represent the
  observed BMSPs listed in Table~\ref{bmsp} and the arrows indicate
  the uncertainties of the observed white dwarf masses. The other
  symbols have the same meaning as in Fig.~\ref{presn}.}
\label{pmwd}
\end{figure}

The parameter space occupied by the BMSPs in the $\log P_{\rm orb}$ -
$M_{\rm WD}$ plane, where $M_{\rm WD}$ is the mass of the white dwarf,
at the time of their formation is displayed in Fig.~\ref{pmwd}. The
positions of the observed BMSPs listed in Table~\ref{bmsp} are
indicated by diamonds. The arrows indicate the uncertainties on the
observed white dwarf masses. The orbital periods of the BMSPs
containing helium white dwarfs are correlated with the masses of the
white dwarfs. The correlation occurs for both the evolutionary
channels He~N and He~T, and arises due to the relation between the
core mass and the radius of the giant which is the progenitor of the
white dwarf, and due to the relation between the Roche-lobe radius of
the giant and the semi-major axis of the orbit (e.g., Rappaport et
al. 1995, Tauris \& Savonije 1999). The BMSPs forming through the
evolutionary channel CO~T are located below the systems
containing helium white dwarfs, while the ones forming through the
direct supernova mechanism CO~N form an extension of the $P_{\rm orb}$
- $M_{\rm WD}$ relation for systems containing a helium white dwarf. 

We conclude this section by noting that the ranges of orbital periods and
secondary masses occupied by the BMSP progenitors at the onset of
Roche-lobe overflow after the supernova explosion of the primary and
at the formation of the BMSP found from our population synthesis code
are in good agreement with those marked out by Tauris et al. (2000) 
and Podsiadlowski et al. (2002) on the basis of detailed numerical
calculations.

\section{Population synthesis}

\subsection{Model assumptions}

Now that we have established the different formation channels
leading to the birth of wide BMSPs and the parameter space occupied by
their progenitors, we turn our attention to the distribution of the
final BMSP orbital periods. In particular, we investigate how
the distribution changes with the assumptions made in the binary
evolution calculations. We therefore repeated the calculations
discussed in the previous sections for a variety of models consisting
of different input parameters. Our reference model, model~A,
corresponds to the input parameters adopted in Sects.~3 and~4. In
models CE1 and CE2, the common-envelope efficiency parameter $\alpha$
is varied.  The role of supernova kicks is explored in models K0,
KM50, KM100, and KM400: in model K0 no kicks are imparted to the
neutron star at birth, while in models KM50, KM100, and KM400, the
Maxwellian velocity dispersion takes the values 50, 100, and
400\,km/s, respectively. In model NS2, the maximum amount of mass that
can be accreted by a neutron star is increased from $0.2\,M_\odot$ to
$0.6\,M_\odot$. The influence of the critical mass ratio $q_{\rm
c,ddi}$ for the development of a delayed dynamical instability is
investigated in models QCDD and QCDD2. In model~C, finally, the
limiting case of conservative mass transfer is considered. Details of
the differences between the various models are summarised in
Table~\ref{models}.

\begin{table*}
\caption{Population synthesis model parameters. }
\label{models}
\begin{tabular}{lcccccc}
\hline
model & $\alpha$ & kicks & $\sigma$ (km/s) & $\gamma_{\rm NS}$ & $q_{\rm c,ddi}$ \\
\hline
A     & 1.0  & yes & 190 & $1-0.2/M_2$ & Hjellming (1989) \\
CE1   & 0.2  & yes & 190 & $1-0.2/M_2$ & Hjellming (1989) \\
CE2   & 5.0  & yes & 190 & $1-0.2/M_2$ & Hjellming (1989) \\
K0    & 1.0  & no  &   - & $1-0.2/M_2$ & Hjellming (1989) \\
KM50  & 1.0  & yes & 50  & $1-0.2/M_2$ & Hjellming (1989) \\
KM100 & 1.0  & yes & 100 & $1-0.2/M_2$ & Hjellming (1989) \\
KM400 & 1.0  & yes & 400 & $1-0.2/M_2$ & Hjellming (1989) \\
NS2   & 1.0  & yes & 190 & $1-0.6/M_2$ & Hjellming (1989) \\
QCDD  & 1.0  & yes & 190 & $1-0.2/M_2$ & 2.5 \\
QCDD2 & 1.0  & yes & 190 & $1-0.2/M_2$ & 3.5 \\
C     & 1.0  & yes & 190 & 0           & Hjellming (1989) \\
\hline
\end{tabular}
\end{table*}

Each time the evolution of an initial binary leads to the formation of
a wide BMSP, its contribution to the orbital period distribution is
weighted according to the probability density distributions of its
initial parameters. The initial mass of the primary (the initially
most massive star) is assumed to be distributed according to the
normalised initial mass function
\begin{equation}
\renewcommand{\arraystretch}{1.4}
\xi \left(M_1 \right) = \left\{
  \begin{array}{ll}
  0 & \hspace{0.3cm} M_1/M_\odot < 0.1, \\
  0.38415\, M_1^{-1} & \hspace{0.3cm} 0.1 \le M_1/M_\odot < 0.75, \\
  0.23556\, M_1^{-2.7} & \hspace{0.3cm} 0.75 \le M_1/M_\odot < \infty,
  \end{array}
\right. \label{imf}
\end{equation}
so that the probability that a star has a mass between $M_1$ and $M_1
+ dM_1$ is given by $\xi \left(M_1 \right) dM_1$. This distribution
function is a simplified version of the initial mass function used by
Hurley et al. (2002) and is based on the initial mass function derived
by Kroupa, Tout \& Gilmore (1993). We recall that for our purpose,
we limit ourselves to masses between $0.1\,M_\odot$ and $60\,M_\odot$.

For a given mass of the primary, the initial mass of the secondary
(the initially least massive star) is determined by the mass ratio
$q=M_2/M_1$. We assume the latter to be distributed according to
\begin{equation}
\renewcommand{\arraystretch}{1.4} n(q) = \left\{
  \begin{array}{ll}
  1 & \hspace{0.3cm} 0 < q \le 1, \\ 0 & \hspace{0.3cm} q > 1,
  \end{array}
\right. \label{imrd}
\end{equation}
so that the masses of the component stars are correlated with each
other. The existence of such a correlation was suggested, among
others, by Eggleton, Fitchett \& Tout (1989).

The distribution of the initial orbital separations is assumed to be
uniform in the logarithm of the semi-major axis $a$. Following Hurley
et al. (2002), we adopt the normalised distribution
\begin{equation}
\renewcommand{\arraystretch}{1.4} \chi (a) = \left\{
  \begin{array}{ll}
  0 & a/R_\odot < 3 \mbox{ or } a/R_\odot > 10^4, \\ 0.12328\, a^{-1}
& 3 \le a/R_\odot \le 10^4.
  \end{array}
\right. \label{iosd}
\end{equation}

After weighting the contribution of each newly formed BMSP to the
total population, the resulting distribution function is convolved
with a constant star-formation rate, on the assumption that the BMSPs
under consideration do not evolve any further once they are formed. In
particular, we presently do not take into account the spin-down time
and thus the finite lifetime of the recycled pulsars. After the
convolution, the distribution function is normalised so that the
integral over all systems found is equal to one.

\subsection{The orbital period distribution}
\label{porb}

\begin{figure*}
\resizebox{7.4cm}{!}{\includegraphics{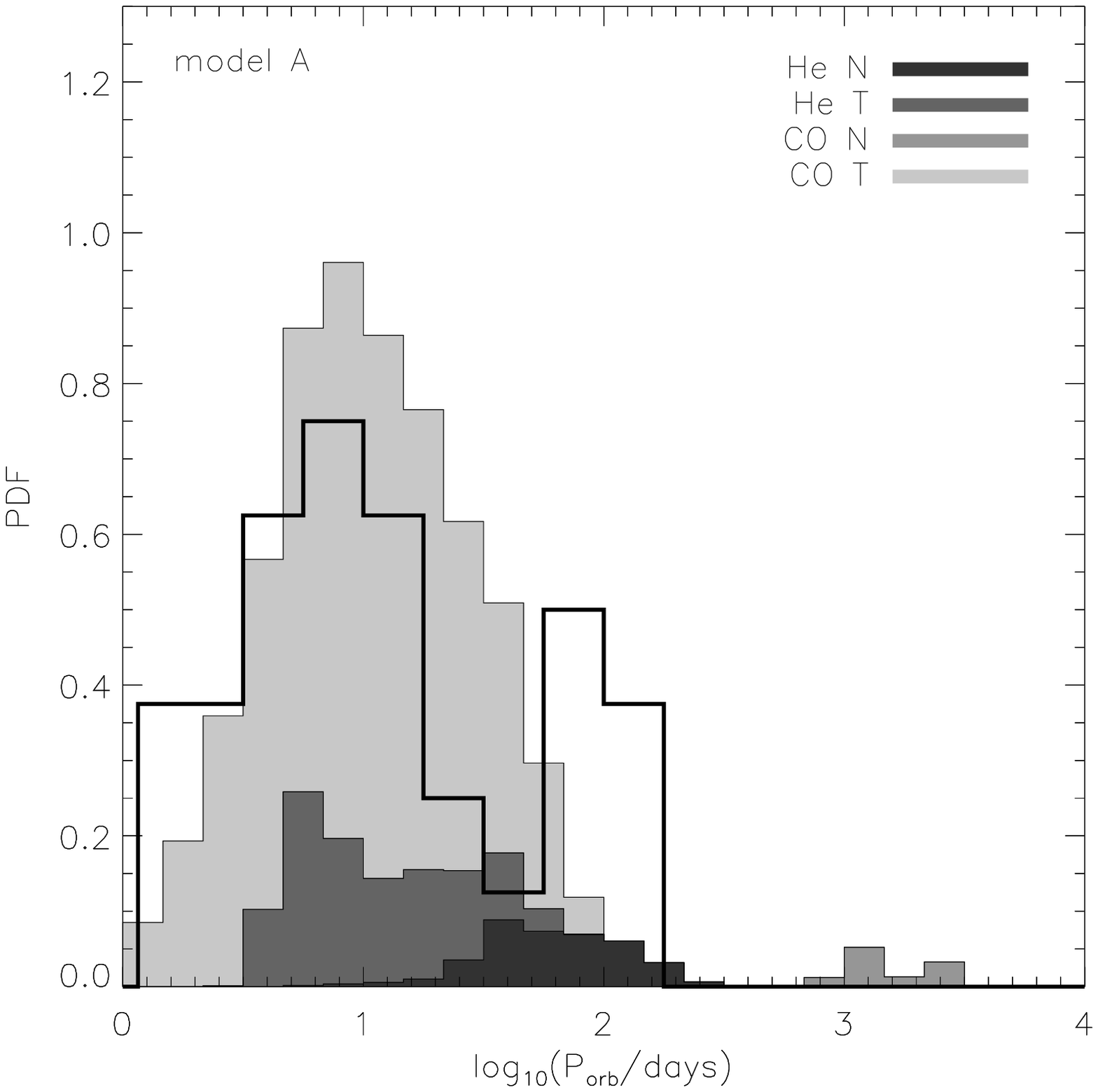}}
\resizebox{7.4cm}{!}{\includegraphics{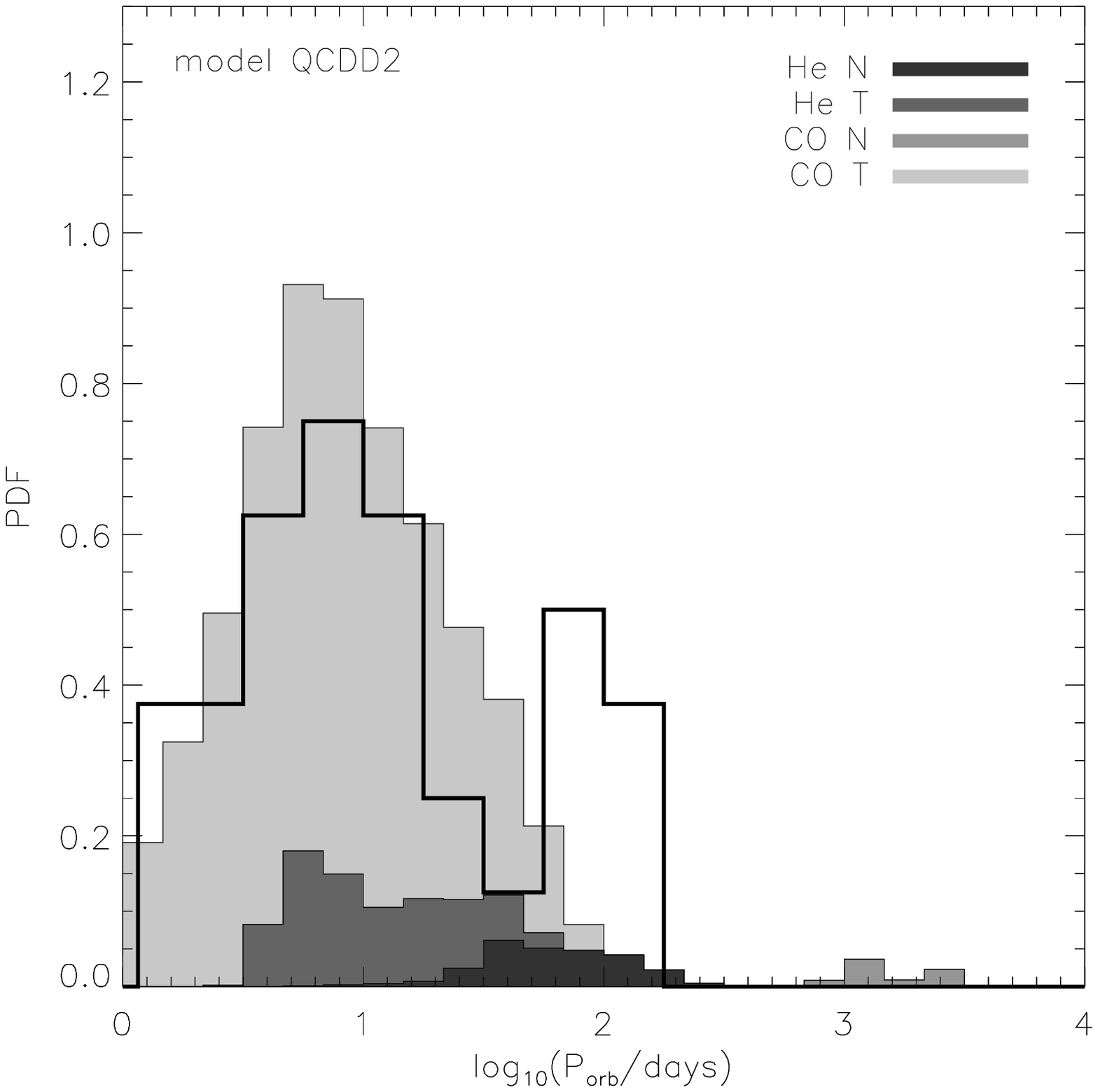}}
\resizebox{7.4cm}{!}{\includegraphics{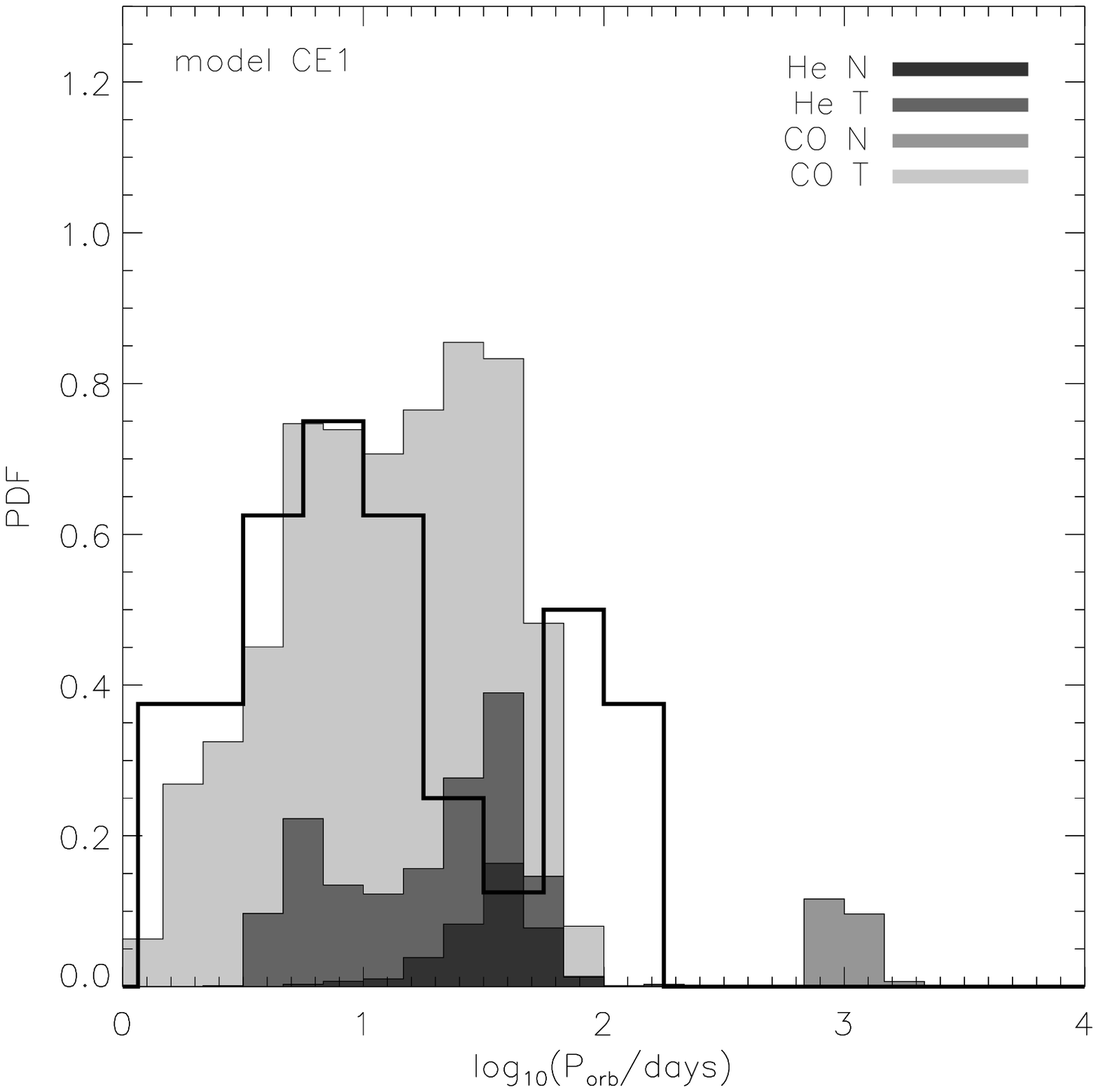}}
\resizebox{7.4cm}{!}{\includegraphics{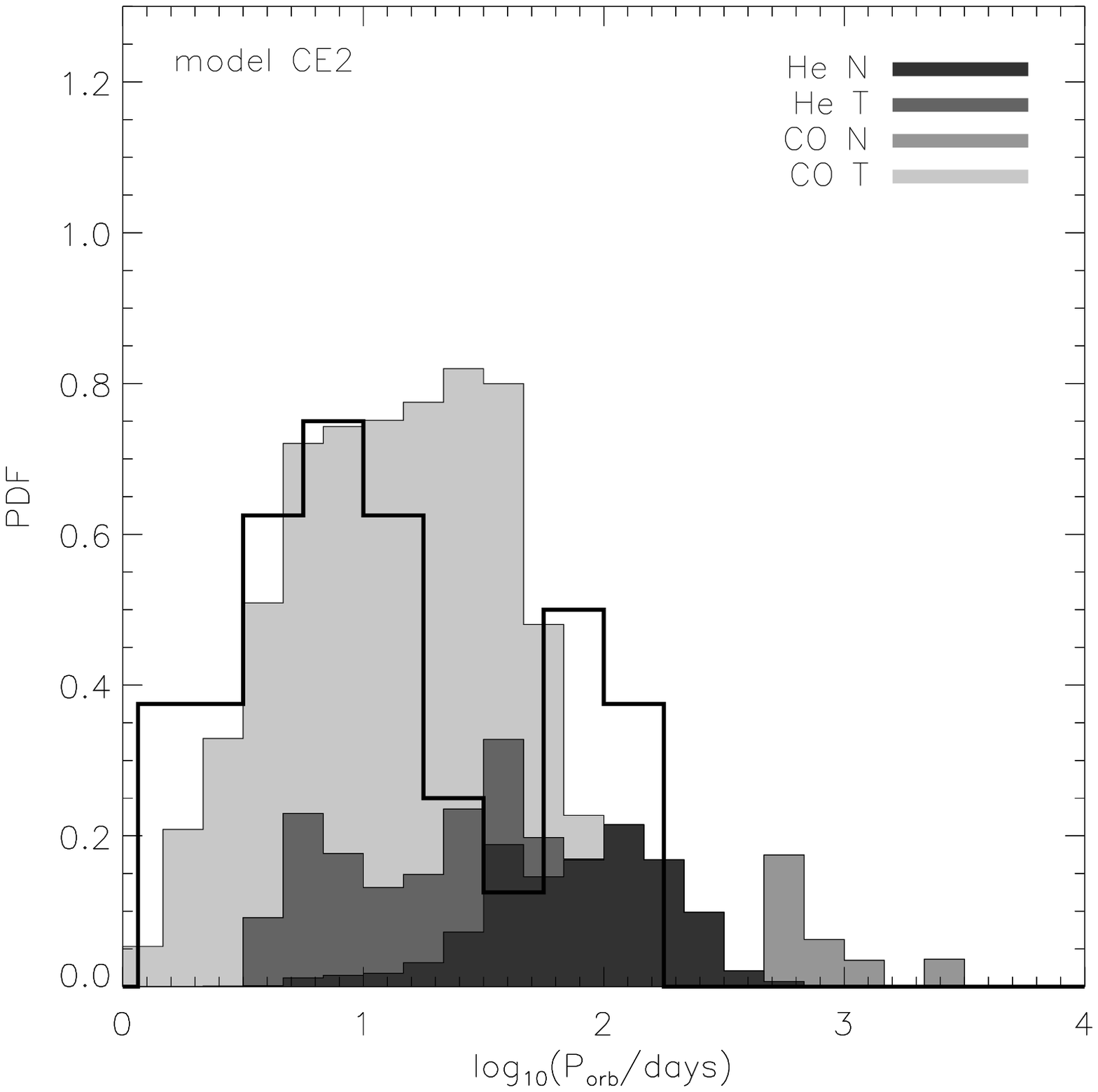}}
\resizebox{7.4cm}{!}{\includegraphics{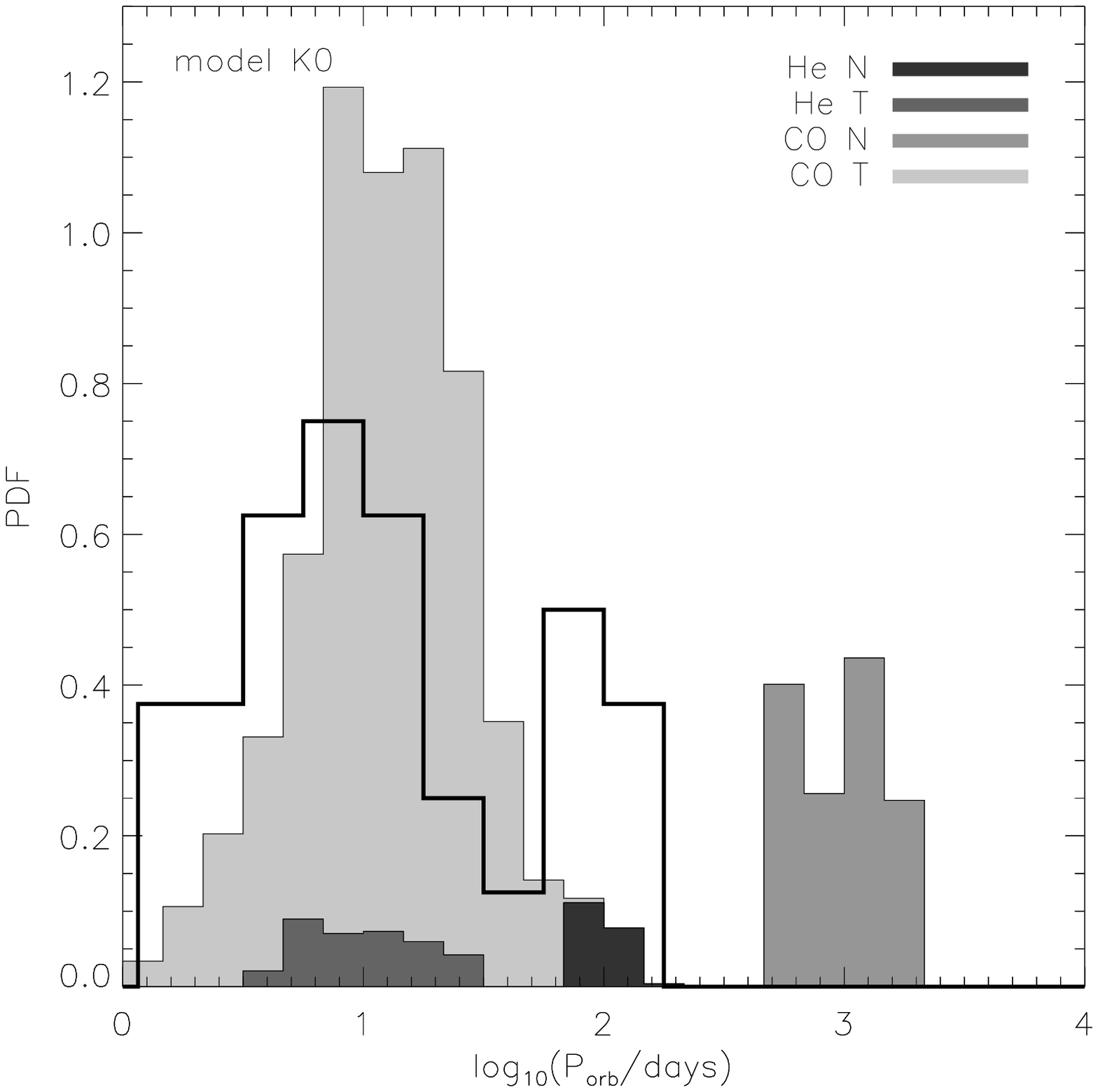}}
\resizebox{7.4cm}{!}{\includegraphics{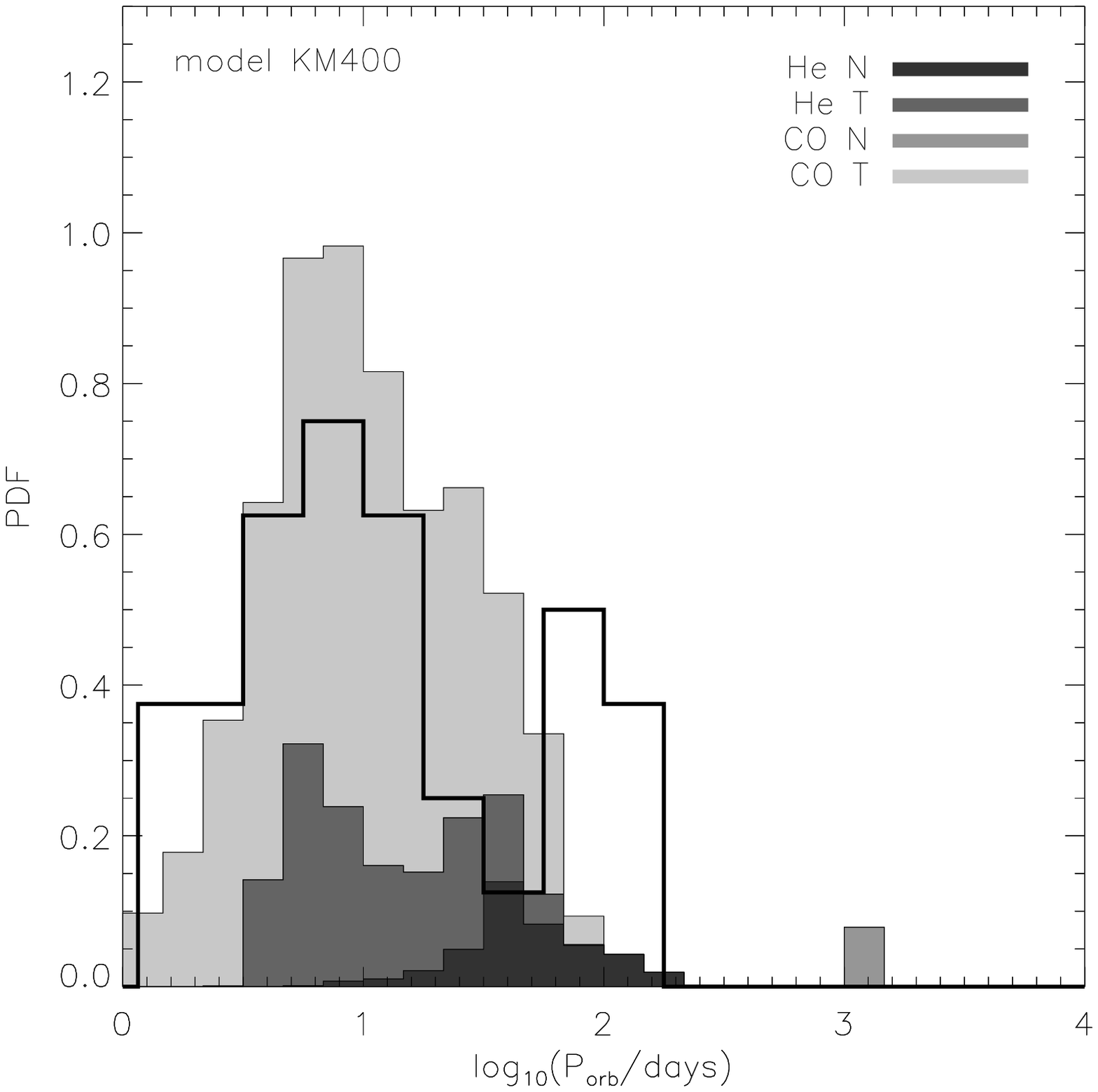}}
\caption{The orbital period distribution of the simulated samples of
  wide BMSPs for the four formation channels considered. The
  distribution functions of the BMSPs forming through the He~N channel
  are multiplied by a factor of~5, and those of the BMSPs forming
  through the CO~N channel by a factor of~100. The black solid line
  represents the orbital period distribution of the observed BMSPs
  listed in Table~\ref{bmsp}.}
\label{stats}
\end{figure*}

We considered the distribution of the orbital periods separately for each
of the evolutionary channels discussed in Sect.~3. The most illustrative
results are shown in Fig.~\ref{stats}, where the contribution of each
channel to the orbital period distribution is displayed using a different
shade of grey. For display purposes, the distribution functions of the
BMSPs forming through the evolutionary channel He~N are multiplied by a
factor of~5, and those of the BMSPs forming through the evolutionary
channel CO~N by a factor of~100. The orbital period distribution of
the observed BMSPs listed in Table~\ref{bmsp} is represented by the
black solid line.

The orbital period distribution of the BMSPs forming through the He~N
channel (black) shows a peak between $\sim\! 50$ and $\sim\! 100$
days, depending on the envelope-ejection
efficiency adopted for the common-envelope phase of the primary and
on the velocity dispersion adopted for the magnitude of the kick
velocity imparted to the neutron star at birth.

A larger envelope-ejection efficiency implies less spiral-in and thus
a larger orbital separation at the end of the common-envelope
phase. If the binary furthermore survives the subsequent
supernova explosion of
the primary, the circularised post-supernova orbital separation must
be smaller than twice the orbital separation before the supernova
explosion (e.g. Kalogera 1996). Increasing the common-envelope
efficiency therefore shifts the peak in the orbital period
distribution to longer orbital periods. If we identify the peak at
long orbital periods in the observed orbital period distribution with
the distribution resulting from the He~N channel, the best agreement
with the observed distribution is found for moderate to large values
of the envelope-ejection efficiency ($\alpha \approx 1.0-5.0$). An
efficiency parameter larger than unity was also suggested by Tauris \&
Bailes (1996) in an investigation on the origin of the millisecond
pulsar velocities and by Tauris \& Sennels (2000) in a study on the
formation of the binary pulsars PSR\,B2303+46 and
PSR\,J1141-6545. Such a large value of $\alpha$ may be 
in conflict with the principle of energy conservation.

The velocity dispersion adopted for the magnitude of the kick velocity
imparted to the neutron star at birth determines the spread of the
orbital periods after the supernova explosion of the primary. Small
velocity dispersions lead to a narrow distribution of orbital periods
close to the upper limit beyond which the binary is disrupted by the
supernova explosion, while larger velocity dispersions produce a
larger spread with more systems at shorter orbital periods
[see Fig.~7 in Kalogera (1996)]. This tendency becomes particularly
clear if one compares the orbital period distribution in the absence
of kicks (model K0) with the orbital period distribution in the case
of large velocity dispersions (model KM400).
The comparison with the observed long orbital period peak is most
satisfactory in the absence of kicks or when the kick velocity
dispersion is not excessively high ($\sigma \la 100\, {\rm km/s}$).

Increasing the amount of mass that can be accreted by the neutron star
from $0.2\,M_\odot$ to $0.6\,M_\odot$ does not significantly change
the orbital period distribution of the BMSPs. In the limiting case of
conservative mass transfer (not shown), the peak in the distribution
is located around the orbital period of 15 days, which is far too
short in comparison to the position of the long-period peak in the observed
distribution. Systems with longer orbital periods do not appear in the
conservative case because their progenitors have donor stars with
masses below $0.9\,M_\odot$, which evolve too slowly to fill their
Roche lobe within the imposed age limit of 15\,Gyr. Changes in the
critical mass ratio for the delayed dynamical instability, finally, do
not influence the orbital period distribution for systems evolving
through this formation channel because the mass ratio of the progenitor
systems at the onset of Roche-lobe overflow is always well below the
critical mass ratio (see Fig.~\ref{postsn}).

In all models considered, few systems are found with orbital periods
longer than $\sim\! 200$ days. The decrease of the simulated
distribution functions at longer orbital periods is, however, a lot
less sharp than the decrease in the observed orbital period
distribution, except for model~K0. The paucity of systems beyond
$\sim\! 200$ days also occurred in a test simulation where we allowed
the neutron star to accrete at the Eddington rate without imposing an
upper limit on the amount of mass that can be accreted. The lack of
long period systems in the simulated orbital period distributions is
therefore not due to the inefficiency of the accretion process
resulting from, e.g., disc instabilities as proposed by Ritter \&
King (2002), but due to the upper limit on the post-supernova orbital
periods which in turn is caused by the upper limit on the initial
orbital periods (see Figs.~\ref{presn} and~\ref{postsn} and the
associated discussion in Sect.~4). The upper limit is preserved by the
mass-transfer process after the supernova explosion due to the limited
range of donor star masses ($0.9\,M_\odot \la M2 \la 1.3\,M_\odot$)
available for stable nuclear timescale mass transfer. This is 
consistent with the results presented by Podsiadlowski, Rappaport \&
Pfahl (2001).

The orbital period distribution of the BMSPs descending from the
evolutionary channel He~T (dark grey) typically has a peak at orbital
periods smaller than 10 days, followed by a long-period tail which can
extend up to $\sim\! 60$ days. The position of the peak is very
robust with regard to changes in the common-envelope efficiency and
the kick velocity dispersion, except in the case of the low envelope
ejection efficiency $\alpha = 0.2$ when the simulated distribution has
a significant secondary peak located in the period gap of the observed
orbital period distribution.

Variations in the assumptions regarding the mass-accretion rate onto
the neutron star again do not significantly influence the
distribution, except in the limiting case of conservative mass
transfer (not shown). In the latter case, mass accretion from donor
stars more massive than $\sim\! 1.8\,M_\odot$ pushes the mass of the
neutron star over the imposed upper limit of $3.0\,M_\odot$, causing
the neutron star to collapse into a black hole. Since mass transfer
for systems evolving through this formation channel initially takes
place on a thermal timescale, the orbital period of the system at the
end of the Roche-lobe overflow phase increases with decreasing mass of
the donor star at the onset of Roche-lobe overflow (e.g. King \&
Ritter 1999). Hence, the lack of systems with initial donor star
masses above $\sim\! 1.8\,M_\odot$ in model~C shifts the peak in the
orbital period distribution towards orbital periods that are too long
in comparison to the position of the short orbital period peak in the
observed distribution. A similar shift takes place when
the critical mass ratio for the delayed dynamical instability is
decreased from the values tabulated by Hjellming (1989) to $q_{\rm
c,ddi}=2.5$ (model QCDD, not shown).

The BMSPs forming through the evolutionary channel CO~T (light grey)
generally dominate the simulated orbital period distribution at short
orbital periods. The distribution function shows a peak around
$\sim\! 10$ days, and a cut-off at $\sim\! 100$ days. In the
conservative mass-transfer model (not shown), neutron stars with
Roche-lobe filling companions more massive than $\sim\!
1.8\,M_\odot$ again collapse into a black hole, so that no BMSPs are
formed through this channel (see also Fig.~\ref{postsn}). A decrease
of the critical mass ratio for the delayed dynamical instability to
$q_{\rm c,ddi}=2.5$ shifts the peak of the distribution function
towards the period gap of the observed orbital period
distribution. All other model parameters do not significantly affect
the orbital period distribution of the BMSPs forming through this
channel.

Comparison of the simulated orbital period distributions resulting
from the formation channels He~N and CO~T shows that for comparable
post-supernova orbital periods, case~B Roche-lobe overflow from a
low-mass donor star can lead to very different final orbital periods
than case~B Roche-lobe overflow from an intermediate-mass donor
star. This behaviour was put forward by Taam et al. (2000) as a
possible explanation for the period gap in the orbital period
distribution of the observed Galactic BMSPs. We note that we here also
find a contribution to the distribution function below the period gap
from systems with intermediate-mass donors evolving through an
initial thermal-timescale case~A mass-transfer phase followed by
a stable case~A/case~B Roche-lobe overflow phase.

Systems evolving into BMSPs through the formation channel CO~N (medium
grey) typically have orbital periods around $\sim\!  1000$ days and
provide the smallest contribution to the total number of BMSPs in our
simulated samples. The contribution of these systems to the total
population of wide BMSPs is largest for kick velocities with a
dispersion of $\sigma \sim 50$\,km/s. It is however clear
that the orbital periods of these BMSPs are too long to contribute to
the orbital period distribution of the observed Galactic BMSPs listed
in Table~\ref{bmsp}. We therefore did not consider these
systems in more detail for the present investigation.

\subsection{Relative contributions}

The relative contributions of the four formation channels to the total
population of wide BMSPs are given in Table~\ref{frac} for each of the
models considered in our investigation. For comparison, the relative
contributions to the observed orbital period distribution of systems
below and above the period gap are about 75\% and 25\%,
respectively.

The simulated populations of wide BMSPs are dominated by the systems
forming through the evolutionary channel CO~T, except when mass
transfer is treated conservatively. However, due to the high
mass-transfer rates inherent to donor stars located in the Hertzsprung
gap, the question may be raised whether neutron stars in binaries
evolving through this evolutionary channel can accrete enough matter
to be spun up to a millisecond pulsar or not. If we impose a lower
limit of $0.05\,M_\odot$ on the mass that needs to be accreted by the
neutron star (see, e.g., Burderi et al. 1999), only binaries with
donor stars initially less massive than $\sim 2.8\,M_\odot$ end up as
BMSPs.  The contributions of the four formation channels to the
population of wide BMSPs then change as indicated in
Table~\ref{frac_0.05}, so that the dominant contribution now stems
from the systems evolving through the evolutionary channel He~T. The
contribution of the systems forming through the evolutionary channel
CO~T decreases even further if a lower limit of $0.1\,M_\odot$ is
imposed on the mass that needs to be accreted by the neutron star. In
this case, the mass of the donor star at the onset of Roche-lobe
overflow must be smaller than about $2.4\,M_\odot$ for the
binary to end up as a BMSP.

\begin{table}
\caption{Relative contributions of the four formation channels
  considered to the total population of wide BMSPs in the
  case of the initial mass ratio distribution given by Eq.\
  (\ref{imrd}).}
\label{frac}
\begin{tabular}{lcccc}
\hline
model & He~N  & He~T  & CO~T  & CO~N \\
\hline
A     &  1.29\% & 17.90\% & 80.79\% &  0.02\%  \\
CE1   &  1.33\% & 19.42\% & 79.21\% &  0.04\%  \\
CE2   &  3.87\% & 17.63\% & 78.46\% &  0.05\%  \\
K0    &  0.65\% &  5.93\% & 93.20\% &  0.22\%  \\
KM50  &  0.23\% &  3.10\% & 96.28\% &  0.39\%  \\
KM100 &  0.53\% &  6.55\% & 92.86\% &  0.05\%  \\
KM400 &  1.43\% & 21.78\% & 76.78\% &  0.01\%  \\
NS2   &  0.79\% & 15.80\% & 83.40\% &  0.01\%  \\
QCDD  &  6.24\% & 23.10\% & 70.56\% &  0.10\%  \\
QCDD2 &  0.90\% & 13.21\% & 85.88\% &  0.01\%  \\
C     & 32.02\% & 67.37\% &  0.00\% &  0.61\%  \\
\hline
\end{tabular}
\end{table}

\begin{table}
\caption{As Table~\ref{frac}, but with a lower limit $\Delta M_{\rm
  NS} = 0.05\,M_\odot$ imposed on the mass that needs to be accreted
  by the neutron star in order to be spun up to a millisecond pulsar.}
\label{frac_0.05}
\begin{tabular}{lcccc}
\hline
model & He~N  & He~T  & CO~T  & CO~N  \\
\hline
A     &  5.64\% & 78.08\% & 16.28\% &  0.01\%  \\
CE1   &  4.52\% & 65.77\% & 29.70\% &  0.02\%  \\
CE2   & 14.50\% & 66.05\% & 19.31\% &  0.14\%  \\
K0    &  9.59\% & 87.87\% &  1.54\% &  0.99\%  \\
KM50  &  6.48\% & 87.72\% &  2.39\% &  3.40\%  \\
KM100 &  6.24\% & 76.82\% & 16.70\% &  0.24\%  \\
KM400 &  5.22\% & 79.31\% & 15.47\% &  0.00\%  \\
NS2   &  3.81\% & 76.12\% & 20.05\% &  0.01\%  \\
QCDD  & 13.25\% & 49.05\% & 37.68\% &  0.02\%  \\
QCDD2 &  5.37\% & 79.11\% & 15.51\% &  0.01\%  \\
C     & 32.03\% & 67.40\% &  0.00\% &  0.57\%  \\
\hline
\end{tabular}
\end{table}

The relative contributions of the four formation channels also depend
on the distribution adopted for the mass ratio of the binaries at the
start of their evolution. Increasing the weight of binaries with
smaller initial mass ratios results in a larger fraction of systems
evolving through the evolutionary channel He~N, so that more BMSPs are
formed with longer orbital periods.  Increasing the weight of binaries
with larger initial mass ratios has the opposite effect.  For
illustration, the relative contributions in the case of the mass ratio
distribution $n(q)=1/q$, for $0 < q \le 1$, are given in
Table~\ref{frac_1q}. In this case, the relative contributions of the
systems evolving through the evolutionary channel He~N approximately
increase by a factor of 3. The overall behaviour of the orbital period
distributions associated with the four formation channels is similar
to that described in Sect.~\ref{porb}.

\begin{table}
\caption{As Table~\ref{frac}, but with an initial mass ratio
  distribution $n(q)=1/q$, for $0 < q \le 1$.}
\label{frac_1q}
\begin{tabular}{lcccc}
\hline
model & He~N  & He~T & CO~T  & CO~N \\
\hline
A     &  4.41\% & 18.72\% & 76.84\% &  0.04\%  \\
CE1   &  3.77\% & 21.46\% & 74.71\% &  0.07\%  \\
CE2   &  9.65\% & 18.99\% & 71.24\% &  0.12\%  \\
K0    &  2.16\% &  6.29\% & 91.02\% &  0.53\%  \\
KM50  &  0.80\% &  3.55\% & 94.65\% &  1.01\%  \\
KM100 &  2.11\% &  7.05\% & 90.71\% &  0.13\%  \\
KM400 &  4.95\% & 24.34\% & 70.69\% &  0.03\%  \\
NS2   &  2.74\% & 15.82\% & 81.41\% &  0.03\%  \\
QCDD  & 14.37\% & 22.57\% & 62.92\% &  0.14\%  \\
QCDD2 &  3.26\% & 14.63\% & 82.08\% &  0.03\%  \\
C     & 37.39\% & 62.06\% &  0.00\% &  0.55\%  \\
\hline
\end{tabular}
\end{table}

As a test, we also considered the case where the mass of the secondary
is weighted independently from the mass of the primary according to the
initial mass function given by Eq. (\ref{imf}). For the models
considered, the corresponding orbital period distributions only show a
peak at short orbital periods if no kicks are imparted to the neutron
star at birth or if the velocity dispersion of the Maxwellian kicks is
small ($\sigma \la 50$\,km/s). However, as can be seen from
Table~\ref{frac_imf}, the contribution of the systems forming
through the evolutionary channel CO~N is then significantly too large
in comparison to the observed orbital period distribution, so that an
independent choice of the mass of both the primary and the secondary
from the adopted initial mass function seems unlikely in this context.

\begin{table}
\caption{As Table~\ref{frac}, but with the initial mass of both the
  primary and the secondary weighted independently according to the
  initial mass function given by Eq. (\ref{imf}).}
\label{frac_imf}
\begin{tabular}{lcccc}
\hline
model & He~N    & He~T  & CO~T  & CO~N  \\
\hline
A     & 21.73\% & 18.25\% & 59.85\% &  0.18\%  \\
CE1   & 15.39\% & 24.15\% & 60.19\% &  0.26\%  \\
CE2   & 37.29\% & 16.56\% & 45.72\% &  0.43\%  \\
K0    & 12.60\% &  5.53\% & 78.12\% &  3.75\%  \\
KM50  &  5.19\% &  3.48\% & 85.54\% &  5.80\%  \\
KM100 & 13.10\% &  6.88\% & 79.32\% &  0.70\%  \\
KM400 & 22.36\% & 24.24\% & 53.24\% &  0.16\%  \\
NS2   & 15.24\% & 15.07\% & 69.54\% &  0.15\%  \\
QCDD  & 41.38\% & 18.62\% & 39.61\% &  0.39\%  \\
QCDD2 & 18.12\% & 15.87\% & 65.86\% &  0.15\%  \\
C     & 47.73\% & 51.45\% &  0.00\% &  0.81\%  \\
\hline
\end{tabular}
\end{table}

\subsection{A 'best-fit' solution?}

\begin{figure}
\resizebox{7.4cm}{!}{\includegraphics{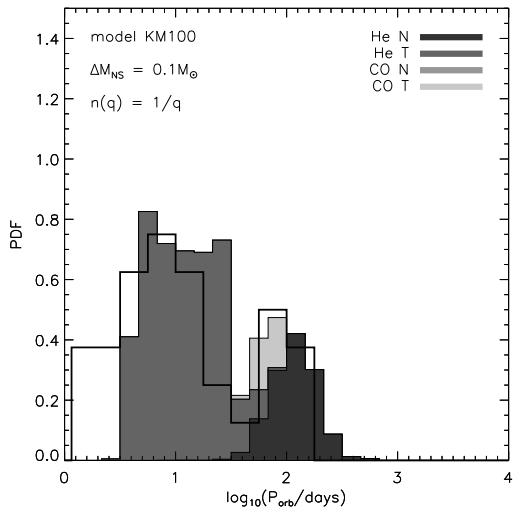}}
\caption{The simulated orbital period distribution for model~KM100
  combined with a lower limit $\Delta M_{\rm NS} = 0.1\,M_\odot$ for
  the mass that needs to be accreted to spin the neutron star up to
  millisecond periods and a mass ratio distribution $n(q)=1/q$, for $0
  < q \le 1$. The black solid line has the same meaning as in
  Fig.~\ref{stats}.}
\label{best}
\end{figure}

From the results presented in the previous subsections, it is clear
that there are still too many uncertainties in some phases of stellar
and binary evolution to come up with a unique set of input parameters
giving the best possible representation of the observed orbital period
distribution. An example of a possible combination of input parameters
reproducing the main features of the observed distribution is shown in
Fig.~\ref{best}. The simulated distribution is obtained from combining
model~KM100 with a lower limit $\Delta M_{\rm NS} = 0.1\,M_\odot$ for
the mass that needs to be accreted to spin the neutron star up to
millisecond periods and a mass ratio distribution $n(q)=1/q$, for $0 <
q \le 1$. Since systems forming through the CO~N channel only accrete
about $0.05\,M_\odot$, they here do not meet the criteria to be
classified as BMSPs.

The combination of parameters reproduces the observed short- and
long-period peaks as well as the observed period gap, but it
fails to reproduce the BMSPs with orbital periods between 1 and 3
days. A possible reason is that the progenitors of these systems 
are likely to converge during the post-supernova mass-transfer phase,
so that they evolve through a different evolutionary channel than
those discussed here.  The model furthermore only predicts BMSPs
containing carbon/oxygen white dwarfs at orbital periods above the
period gap, which may pose a problem to explain some of the observed
systems with more massive white dwarfs and orbital periods shorter
than 20 days (see Fig.~\ref{pmwd}). This potential inconsistency
could be resolved if the uncertainties in the accretion process onto
neutron stars during the rapid thermal-timescale mass-transfer phase
of the systems evolving through the CO~T channel is better
understood.

The relative contributions of the systems forming through the He~T,
He~N, and CO~T channels are 72.51\%, 21.65\%, and 5.84\%,
respectively, which agrees fairly well with the relative number
of systems below and above the period gap in the observed orbital
period distribution. The agreement is even better if the observed
distribution is restricted to orbital periods longer than 3 days, in
which case 70\% of the observed systems have orbital periods below the
gap and 30\% have orbital periods above the gap.

\section{Concluding remarks}

We have investigated the formation of wide BMSPs through four
evolutionary channels using a rapid binary evolution code based on the
analytical approximation formulae for the evolution of single stars
derived by Hurley et al. (2000).

In three of the channels, dynamically unstable mass transfer from the
primary prior to the supernova explosion results in a common-envelope
phase during which the orbital separation is reduced and the helium
core of the primary is exposed as a naked helium star. The
primary continues its evolution until it becomes a neutron star in a
type Ib/c supernova explosion. The further evolution depends on
the mass of the secondary and on the post-supernova orbital period
after circularisation:

\begin{itemize}
\item If the mass of the secondary is smaller than $\sim\! 1.6\,
  M_\odot$, the binary undergoes a stable case~B mass-transfer phase
  when the secondary reaches the giant branch. The binary becomes a
  BMSP consisting of a rapidly rotating neutron star orbiting a helium
  white dwarf with an orbital period determined by the mass of the
  white dwarf.
\item If the mass of the secondary is larger than $\sim\! 1.6\,
  M_\odot$ and the orbital period shorter than $\sim\! 2.5$ days,
  the binary undergoes a thermal-timescale case~A mass-transfer phase,
  followed by a stable phase of case~A and/or case~B Roche-lobe
  overflow. The system again evolves into a BMSP containing a helium
  white dwarf whose mass determines the orbital period.
\item If the mass of the secondary is larger than $\sim\! 1.6\,
  M_\odot$ and the orbital period longer than $\sim\! 2.5$ days,
  the binary undergoes a thermal-timescale case~B mass-transfer phase
  when the secondary crosses the Hertzsprung gap. The
  system evolves into a binary containing a neutron star and a
  carbon/oxygen white dwarf without developing a tight correlation
  between the orbital period and the mass of the white dwarf. The
  accretion onto the neutron star in these binaries may not be
  efficient enough to spin them up to millisecond pulsars.
\end{itemize}

The fourth evolutionary channel corresponds to the direct supernova
mechanism where no interaction between the binary components takes
place until the primary gives birth to a neutron star in a type II
supernova explosion on the AGB. If the resulting
post-supernova orbit is wide enough to avoid any interaction until the
secondary becomes an AGB star, stable case~C mass
transfer leads to the formation of a BMSP containing a carbon/oxygen
white dwarf whose mass is correlated with the orbital period.

Our main goal in this investigation was to compare the orbital
period distribution of
simulated samples of BMSPs resulting from different sets of input
parameters for the binary evolution calculations with the orbital
period distribution of wide BMSPs observed in the Galactic disc. The
observed long orbital period peak is identified with the BMSPs
containing a helium white dwarf whose progenitor underwent a
mass-transfer phase that was stable at all times. The observed short
orbital period peak on the other hand is identified with the BMSPs
containing a helium white dwarf whose progenitor evolved through a
thermal-timescale mass-transfer phase on the main sequence or in the
Hertzsprung gap.

The simulated distribution functions all show a rapid decrease for
orbital periods longer than 200 days, irrespective of the accretion
efficiency of neutron stars. The lack of longer period systems is a
consequence of the upper limit on the initial orbital periods beyond
which the binary remains detached instead of going through a
common-envelope phase prior to the supernova explosion of the
primary. The agreement between the simulated and the observed orbital
period distributions is best for models with highly non-conservative
mass transfer, common-envelope efficiencies equal to or larger than
unity, a critical mass ratio for the delayed dynamical instability
larger than 3, and no or moderate supernova kicks at the birth of the
neutron star.

The first results of the population synthesis study presented here are
encouraging, but many uncertainties remain to be solved. The lack of
BMSPs containing carbon/oxygen white dwarfs at orbital periods shorter
than 20 days in our 'best-fit' solution, for instance, surely poses a
problem that needs to be resolved. A more detailed treatment of the
accretion process onto neutron stars and the potential effects of
pulsar turn-on, evaporation, and spin-down are of particular
importance to improve the population synthesis study of wide
BMSPs. Preliminary calculations also reveal a dependency of the
orbital period distribution on the mass-loss rates from stellar winds,
but the results are still inconclusive.

Furthermore, our population synthesis study implicitly
incorporates a model for the Galactic population of neutron star X-ray
binaries. The long-standing apparent conflict between the
observationally implied birth rate of BMSPs and their LMXB progenitors
(e.g. Ruderman, Shaham \& Tavani 1989; Podsiadlowski et al. 2002 and
references therein) may point towards non-standard evolutionary
effects in the X-ray binary phase which are not considered in our
study. It is therefore desirable to consider the populations of LMXBs
and BMSPs simultaneously.

We will address these matters in more detail in future
investigations.

\section*{Acknowledgements}
We are grateful to Hans Ritter for providing the data of the observed
BMSPs and to Jarrod Hurley, Onno Pols, and Chris Tout for sharing
their SSE software package and a copy of the paper describing their
binary evolution algorithm prior to publication. We also thank Jarrod
Hurley, Chris Tout, Simon Portegies-Zwart, and Firoza Sutaria for
useful discussions. We thank the referee, Philipp Podsiadlowski, for
useful comments. This research was supported by the British Particle
Physics and Astronomy Research Council (PPARC).

\appendix

\section{Accretion onto neutron stars}
\label{rlof}

In this appendix, we briefly outline the treatment of mass accretion
by neutron stars in semi-detached binaries adopted in our binary
evolution code. We assume mass transfer to be non-conservative and
denote by $\gamma$ the fraction of the transferred mass that is lost
from the system. The mass-accretion rate onto the neutron star is then
related to the mass-transfer rate from the donor star by
\begin{equation}
\dot{M}_{\rm NS} = (1-\gamma) |\dot{M}_2|, \label{Ma}
\end{equation}
where $M_{\rm NS}$ is the mass of the neutron star, and $M_2$ the mass
of the Roche-lobe overflowing companion.

If we furthermore assume that the neutron star accretes a maximum
mass $\left(\Delta M_{\rm NS} \right)_{\rm max}$ during the mass
transfer process, the average mass-accretion rate onto the neutron
star is given by
\begin{equation}
\dot{M}_{\rm NS} = \left(\Delta M_{\rm NS}\right)_{\rm max}
  {{|\dot{M}_2|} \over M_2}, \label{Ma2}
\end{equation}
so that
\begin{equation}
\gamma = 1 - {{\left(\Delta M_{\rm NS}\right)_{\rm max}} \over
  M_2}. \label{Ma3}
\end{equation}
In order to ensure that the neutron star accretes not more than
$\left(\Delta M_{\rm NS} \right)_{\rm max}$, we then adopt the
artificial rate given by~(\ref{Ma2}) as the mass-accretion rate onto
the neutron star.

\bsp

\label{lastpage}

\end{document}